\documentclass[twocolumn]{autart}    

\usepackage{graphicx}          
\usepackage{natbib}
\usepackage{amssymb}
\usepackage{multirow}
\usepackage{makecell}
\usepackage{booktabs}
\usepackage{graphicx}
\usepackage{lineno}
\usepackage[cmintegrals]{newtxmath}
\usepackage{bm}
\usepackage{amsmath}
\usepackage{enumerate}
\usepackage{color}
\usepackage{graphics}
\usepackage{multicol}
\newcommand{\ds}{\displaystyle}
\newtheorem{theorem}{Theorem}[section]
\newtheorem{corollary}[theorem]{Corollary}
\newtheorem{definition}[theorem]{Definition}
\newtheorem{problem}[theorem]{Problem}

\newtheorem{remark}[theorem]{Remark}

\newtheorem{assumption}[theorem]{Assumption}
\newtheorem{proposition}[theorem]{Proposition}
\newtheorem{lemma}[theorem]{Lemma}
\newtheorem{example}[theorem]{Example}
\begin{document}

\begin{frontmatter}

\title{Synchronization of Power Systems \\under Stochastic Disturbances} 

\thanks[footnoteinfo]{This paper was not presented at any IFAC 
meeting. This work is financially supported 
by the Foundation for Innovative Research Groups of the National Natural Science 
Foundation of China with grant No. 61821004, the National Natural Science Foundation of China with grant No.  62103235 and the Natural Science Foundation of Shandong Province with grant No. ZR2020QF118. Corresponding author: Kaihua Xi.}

\author[SDU]{Zhen Wang},    
\author[SDU]{Kaihua Xi}\ead{KXI@sdu.edu.cn},               
\author[SDU]{Aijie Cheng},  
\author[TUD,LEIDEN]{Hai Xiang Lin},
\author[AMS,AMS2]{Andr\'{e} C. M. Ran},
\author[TUD]{Jan H. van Schuppen} ,
\author[SDU2]{Chenghui Zhang} 

\address[SDU]{School of Mathematics, Shandong University, Jinan, 250100, P. R. China }  
\address[TUD]{Delft Institute of Applied Mathematics, Delft University of Technology, Delft, 2628 CD, The Netherlands}
\address[LEIDEN]{Institute of Environmental Sciences (CML), Leiden University, Leiden 2333 CC, The Netherlands} 
\address[AMS]{Department of Mathematics, Faculty of Science, Vrije Universiteit, Amsterdam, 1081 HV, The Netherlands} 
\address[AMS2]{Research Focus: Pure and Applied Analytics, North-West University,
Potchefstroom, South Africa }            
\address[SDU2]{School of Control Science and Engineering, Shandong University, Jinan, 250061, P. R. China}        


\begin{abstract}                          
The synchronization of power generators is an important condition for the proper functioning of a power system, in which the fluctuations in frequency and the phase angle differences between the generators are sufficiently small when subjected to stochastic disturbances. Serious fluctuations can prompt desynchronization, which may lead to widespread power outages. 
Here, we model the stochastic disturbance by a Brownian motion process in the linearized system of the non-linear power systems and characterize the fluctuations by the variances of the frequency and the phase angle differences in the invariant probability distribution. We propose a method to 
calculate the variances of the frequency and the phase angle differences. For the system with uniform disturbance-damping ratio, we derive explicit formulas for the variance matrices of the frequency and the phase angle differences. It is shown that the fluctuation of the frequency at a node depends 
on the disturbance-damping ratio and the inertia at this node only, and the fluctuations of 
the phase angle differences in the lines are independent of the inertia. 
In particular, the synchronization stability is related to the cycle space of the network. We reveal the influences of constructing new lines and increasing capacities of lines on the fluctuations in the phase angle differences in the existing lines. The results are
illustrated for the transmission system of Shandong Province of China. For the system with non-uniform disturbance-damping ratio, we further obtain bounds of the variance matrices.
\end{abstract}

\end{frontmatter}

\section{Introduction}
\par 
Power grids deliver a growing share of the energy consumed in the world and are undergoing an unprecedented revolution because of the increasing integration of intermittent power sources such as solar and wind energy and the commercialization of plug-in electric automobiles. These developments will change the structure of power sources and decrease carbon emissions dramatically, but they will also lead to new disturbances associated with fluctuations in energy production and load. These disturbances not only deteriorate the quality of the power supply but may trigger loss of synchronization, which can result in serious blackouts \citep{upgrading_the_grid}. This indicates the necessity to study synchronization under stochastic disturbances.
%
%
%
%

\par 
Here, we focus on the synchronization of power systems under stochastic disturbances. We explore the role of system parameters in a framework of stochastic systems that can be extended to other real complex networks with synchronization. In a synchronous state of a power system, the frequencies of the synchronous machines (e.g., rotor-generators driven by steam or gas turbines) should all be equal or close to the nominal frequency (e.g., 50 Hz or 60 Hz). Here, the frequency is the derivative of the rotational phase angle and is equal to the rotational speed of the synchronous machine in units of $rad/s$. The synchronization stability is defined as the ability to maintain synchronization under disturbances, which is also called transient stability \cite{Kundur}.The parameters that determine synchronization include the power flows, inertia \citep{optimal_inertia_placement} and damping coefficients \citep{motter,NISHIKAWA20151} of the synchronous machines as well as the coupling strength \citep{FAZLYAB2017181} between the synchronous machines and the network topology, which can be optimized to enhance stability by load-frequency control or by constructing new power generators, virtual inertia and  transmission lines. In the analysis of the existence condition of a synchronous state \citep{Dorfler2012} and the linear \citep{motter,NISHIKAWA20151} and nonlinear stability of that state \citep{zab2,menck,hirsch1}, the focus is on the synchronous state, on the local convergence or on the basin of attraction.
However, in practice, the state of the power system never stays at the synchronous state and is always fluctuating due to various disturbances. 
If both the fluctuations of the frequency and the phase angle difference are so large that the system cannot return to the synchronous state, then the synchronization is lost. 
Hence, the impact of the disturbances cannot be neglected and the size of the fluctuations directly characterizes the stability of the system.
\par 
Robust control methods in load frequency control may be used to improve 
the stability, where the disturbances are considered,  see \cite{Robustload,passivityrobust,MLPIAC}. By these methods, the power generation are controlled 
to balance the disturbances. However,  besides the power generations, the stability of the system also depends
on the network topology, line capacities, inertia of the synchronous machines and so on,  for which the values cannot be specified by the robust control methods.  
 By modelling the disturbances as inputs to the associated linearized system, the fluctuations are evaluated by the $\mathcal{H}_2$ norm of the input-output linear system \citep{H2norm,optimal_inertia_placement}. However, because the $\mathcal{H}_2$ norm equals to the trace of a matrix \citep{H2norm_another_form}, which is a global 
metric for the synchronization stability, the fluctuations of the frequency at each node, the phase angle difference in each line and their correlation can hardly be explicitly characterized.  Clearly, the nodes with serious fluctuations in the frequencies and the lines with serious fluctuations in the phase angle differences are vulnerable to disturbances. 
In physics, the focus is on the propagation 
of the disturbances \citep{Propagation1,Delocalization,topological-spreading,Aue17,Zha19} and the network susceptibility \citep{Networksuseptbilities}. For example, 
the statistics of the fluctuations at the nodes, e.g., the variance of the 
increment of the frequency distribution, can be calculated via simulations by modelling the disturbances by either Gaussian or non-Gaussian noise \citep{Propagation1}. With perturbations added to the system parameters, 
the disturbance arrival time and the vertex and edge susceptibility are estimated in \citep{topological-spreading,Networksuseptbilities} respectively.  The amplitude of perturbation responses of the nodes is used to
study the emergent complex response patterns across the network \citep{Zha19}. 
By these investigations on fluctuations, intuitive insights on the impact of the system parameters, e.g., the network topology and the inertia of synchronous machines, on the spread of the disturbances are provided, which may help to develop
practical guiding principles for real network design and control. 
\par 
In this paper, we investigate the fluctuations of the frequency at each node and the phase angle difference in each line in a linear stochastic system. By modelling the disturbances by Gaussian noise, we use the variance in the invariant probability distribution to characterize the fluctuations and  propose an efficient method for the calculation of the variance by solving a Lyapunov equation instead of statistics with a large amount of simulations.  Under assumptions of uniform disturbance-damping ratio at the nodes, explicit formulas for the variances of the fluctuations in the frequencies and phase angle differences are derived, which can be used to tune the system parameters to improve the synchronization stability.  With these explicit formulas, the impact of the network topology on the synchronization stability is considerably clarified.
\par 
The contribution of this paper include:
\begin{enumerate}[(i)]
\item a new metric, that is the variance in the invariant probability distribution of the frequencies at the nodes and 
the phase angle differences in the lines, is proposed for the analysis of the synchronization stability.
With this metric, the vulnerable nodes and lines can be identified effectively based on solving a matrix Lyapunov equation;
\item under the assumption that the disturbance-damping ratio is uniform,  we derive an explicit formula of the variance matrix of the frequency, which reveals the impact of the inertia and the disturbance-damping ratio, and an explicit formula of the variance matrix of the phase angle differences,  which 
reveals the impact of the network topology and the disturbance-damping ratio;
\item for non-uniform disturbance-damping ratio, an upper and a lower bound of the variance matrices
of the  frequency and the phase angle differences are deduced;
\item the impact of constructing new lines and increasing the capacity of lines on the variance 
are investigated. 
\end{enumerate} 
The findings of this paper provide directions for the optimization of the droop control coefficients,
the placement of virtual inertia and energy storage, changes in the network topology,
and changes in the capacity of lines in the power systems. The framework of this paper for the
investigation of synchronization stability may be extended to other networks with stochastic
disturbances and problems of synchronization.

\par 
This paper is organized as follows. 
The mathematical model of the power system  and 
the problem formulation are introduced in Section \ref{Section:Model}.
We propose a method to calculate the variance matrices of the invariant probability distribution 
of the frequency and phase angle differences in Section \ref{Section:calculation}. 
We derive explicit formulas for the two matrices in Section \ref{Section:formula}.
Based on the explicit formulas, we deduce bounds of the variance matrices for the networks 
with non-uniform disturbance-damping ratio in Section \ref{Section:bound}. 
We study the role of the network topology in Section \ref{Section:network} with verification using a 
real network in Section \ref{casestudy}.  We conclude this paper 
with perspectives in Section \ref{Section:conclusion}. 
\par

\section{Models and problem formulation}\label{Section:Model}

The power grid can be modelled by a graph $\mathcal{G}(\mathcal{V},\mathcal{E})$ with nodes
$\mathcal{V}$ and edges $\mathcal{E}\subset \mathcal{V}\times\mathcal{V}$, where a node represents a bus and an edge $(i,j)$ represents the transmission line between nodes $i$ and $j$. We focus on the transmission network and assume the lines are lossless. We denote the number of nodes in $\mathcal{V}$ and edges in $\mathcal{E}$ by $n$ and $m$, respectively. The dynamics 
of the power system are described by the following swing equations (\citep{zab2,menck2,hirsch1}):  
\begin{subequations}\label{Nonlinear}
\begin{align}
\dot{\delta}_i&=\omega_i,\\
m_i\dot{\omega}_i&=P_i-d_i\omega_i-\sum_{j=1}^nl_{ij}\sin{(\delta_i-\delta_j)},
\end{align}
\end{subequations}
where $\delta_i$ and $\omega_i$ denote the phase angle and the frequency deviation of the synchronous machine at node $i$; $m_i>0$ describes the inertia of the synchronous generators; $P_i$ denotes power generation 
if $P_i>0$ and denotes power load otherwise; 
$l_{ij}=\hat{b}_{i j}V_iV_j$ is the effective susceptance, where $V_i$ is the voltage; $d_i>0$ is the damping coefficient with droop control.  Since the dynamics of the voltage and the frequency can be decoupled \citep{Simpson-Porco2016},we restrict attention to modelling only the dynamics of the frequency and assume that the voltage of each node is a constant. In practice, the voltage can be well controlled by an \emph{Automatic Voltage Regulator \citep{Kundur}}.
This model is often applied to study transient stability and rotor angle stability \citep{Nishikawa_2015,menck2,Dorfler2012}.  
In this paper, we focus on the networks with the following assumption.
\begin{assumption}\label{assumption0}
Assume the network $\mathcal{G}(\mathcal{V},\mathcal{E})$ is connected. 
\end{assumption}
With Assumption \ref{assumption0},  we easily obtain $m\geq n-1$. In special case of $m=n-1$, the network is acyclic  and 
when $m\geq n$,  there are cycles in the network.  

\subsection{The synchronous state}
The stable region of system (\ref{Nonlinear}) is analysed 
by \cite{hirsch1} \emph{et al} and \cite{zab2}.  
The stability analysis of a power system makes use of the
concept of the synchronous state which satisfies,
for $i=1,2,\cdots,n$, 
\begin{align*}
\omega_i(t)=\omega_{syn},~\text{and}~
\delta_i(t)=\omega_{syn}t+\delta_i^*,
\end{align*}
where $\delta_i^*$ and the synchronized frequency $\omega_{syn}$ satisfy,
\begin{subequations}
\begin{align*}
P_i-D_i\omega_{syn}-\sum_{j=1}^nl_{ij}\sin{\delta_{ij}^*}&=0~~ \text{for}~~ i=1,\cdots,n,\\
\omega_{syn}-\frac{\sum_{i=1}^{n}{P_i}}{\sum_{i=1}^{n}D_i}&=0,
\end{align*}
\end{subequations}
where $\delta_{ij}^*=\delta_i^*-\delta_j^*$ is the phase angle difference between nodes $i$ and $j$, which are directly connected by transmission line $(i,j)$. 
The power flow in line $(i,j)$ is 
$l_{ij}\sin{\delta_{ij}^*}$, which is determined by the load frequency control \citep{Kundur}.
$\omega_{syn}$ is the deviation of the synchronized frequency from the nominal value of frequency.
There are three forms of load frequency control distinguished from fast to slow time-scales, i.e., primary, secondary and tertiary
frequency control.  Primary control maintains the synchronous state by droop control on a small time-scale.
However, this synchronized frequency may deviate
from its nominal value in a medium time-scale, which leads to $\omega_{syn}\neq 0$. 
Secondary control restores the synchronized frequency to its nominal value such that $\omega_{syn}=0$ 
on a medium time-scale. With a prediction
of power demand, tertiary control calculates the operating
point stabilized by primary and secondary control on a large
time-scale, which concerns the security and economy of the
power system.  In the control design for
frequency synchronization, the power input $P_i$ is determined 
in the secondary and tertiary control.  Thus, it is practical to assume that the power generation and load are balanced
in the study of frequency synchronization. Thus, $\sum_{i=1}^{n}{P_i}=0$, which leads to $\omega_{syn}=0$. 
\par 
Due to low line capacities, the synchronous state might not exist. For 
the condition of the existence of the synchronous state, we refer to 
\cite{Dorfler20141539}. For the number of the synchronous states, see \cite{luxem,baillieul}. 

\subsection{The linearized model}

Assume that there exists a synchronous state $(\bm\delta^*,\bm 0)$ for system (\ref{Nonlinear}), which  can be linearized as
\begin{align}\label{Linearization}
\left(\begin{array}{c} {\bm{\dot{\delta}}} \\ {\bm{ \dot{\omega}}}\end{array}\right)&=\left(\begin{array}{cc} \bm 0 &\bm I_n\\
-\bm M^{-1}\bm L_c & -\bm M^{-1}\bm D\end{array}\right) \left(\begin{array}{c} {\bm\delta} \\{\bm \omega}\end{array}\right),
\end{align}
where $\bm\delta=\text{col}{(\delta_i)}\in\mathbb{R}^n$, $\bm I_n\in\mathbb{R}^{n\times n}$ is the identity matrix, $\bm\omega=\text{col}{(\omega_i)}\in\mathbb{R}^n$, $\bm M=\text{diag}(m_i)\in\mathbb{R}^{n\times n}$, $\bm D=\text{diag}{(d_i)}\in\mathbb{R}^{n\times n}$, and $\bm L_c=(\widetilde{l}_{c_{ij}})\in\mathbb{R}^{n\times n}$ is the Laplacian matrix of the network with weight $l_{ij}\cos{\delta_{ij}^*}$ generated by $(\bm\delta^*,\bm 0)$, which satisfies
\begin{align}\label{Laplacianform}
 \widetilde{l}_{c_{ij}}=\begin{cases}
        ~~- l_{ij}\cos{\delta_{ij}^*},~~ i\neq j,\\
        ~~\ds  -\sum_{k\neq i}\widetilde{l}_{c_{ik}},~~ i=j.
        \end{cases}
\end{align}
 Note that the state variables in (\ref{Linearization}) are the deviations of the phase angles and frequencies from the synchronous state $(\bm\delta^*,\bm 0)$. By the second Lyapunov method, the stability of $(\bm\delta^*,\bm 0)$ can be 
determined by the sign of the real part of the eigenvalues of the system matrix of (\ref{Linearization}). The analysis 
of the eigenvalue of the system matrix is also called \emph{small-signal stability analysis}.
It has been proven that if $l_{ij}\cos{\delta_{ij}^*}\geq 0$, then the system is stable  at the synchronous state $(\bm \delta^*,\bm 0)$ \citep{zab1}, which leads to the 
security condition 
\begin{align}\label{condition}
\bm\Theta=\big\{\bm\delta\in\mathbb{R}^n \big |~|\delta_{ij}|< \frac{\pi}{2},\forall (i,j)\in \mathcal{E})\big\}.
\end{align}
It has been proven by \citep{skar_uniqueness_equilibrium} that 
for the power network with a general network topology, the synchronous 
state in this security range is unique. For the identification of the subset of the $n$-torus where there exists a synchronous state, we refer to \citep{Jafarpour}. 

\subsection{Problem formulation}

In real networks, the state of the power system always fluctuates around the synchronous state due to various disturbances. 
If the fluctuations are very large,
the state may exit the stability region of the synchronous state and lead to instability of the system. 
A sign of desynchronization is that both the fluctuations of the frequency and the phase angle difference are so large that the system cannot return to the synchronous state. 
Many factors influence the fluctuations, which 
include the parameters of the transmission lines, the synchronous machines,
the network topology and the disturbances. The source of the
 disturbances are also various, e.g., the renewable power generation, fault of the devices in the network, etc. 
We focus on the following
problem in this paper. 
\begin{problem}\label{problem1}
How do the fluctuations of the 
frequency and the phase angle differences depend on the parameters of the system 
and the disturbances?
\end{problem}
The solution of this problem provides insights for suppressing the fluctuations by scientific
 parameter assignments. 
 The choice of a model for the fluctuations in a power system should be based on the criteria that the model is realistic and that the subsequent analysis is not too complex. 
\par 
A realistic model of the actual disturbances affecting a power system at each node requires an extensive system identification procedure, including the collection of a large amount of data on the fluctuations of the power system. The disturbances come from both the loads and the various power sources, such as wind parks and photovoltaic units. 
It has been shown that the probability distributions of disturbances are not Gaussian in several real
power grids, e.g., grids in North America and
Europe, which leads to non-Gaussian 
distribution of the frequency and is crucial to induce desynchronization in the system, see \cite{Propagation1,non-gaussian1,non-gaussian3,non-gaussian2,LXIEGaussian} etc..
A model could then be a nonlinear stochastic differential equation of the power system driven by either Brownian motion or another process with independent increments. However, the performance evaluation of such nonlinear stochastic system requires requires either the numerical approximation of the solution of a partial differential equation \citep{KeyouWang} or
a large amount of simulations for the statistics of the frequencies \citep{Propagation1}. This model is too complicated to obtain an analytic probability distribution of the state of the power system consisting of a large number of synchronous machines.
\par 
An alternative to the modelling approach described above is to formulate a deterministic linear system obtained by linearization of a nonlinear power system at a synchronous state. The deterministic linear system is then transformed into a linear stochastic differential equation driven by Brownian motion. Such models are often used in control engineering and in mathematical finance, and these models are regarded as reasonable approximations of realistic models. Moreover, these models have a low algebraic complexity. 
It is well known that for a linear stochastic differential equation with a system matrix that is Hurwitz, there exists an invariant probability distribution of the state that is a Gaussian probability distribution characterized by the mean value and the variance of the state \cite[Theorem 1.53]{linearOptimalSystems}\cite[Theorem 6.7]{karatzas:shreve:1988}.  For power systems, the fluctuations are described by the variance matrices in the invariant probability distribution of the associated linear stochastic system. The dependence on the system parameters is indicated. The complexity of the performance of this model is manageable. Though the analysis of the stochastic linearized system is valid only for 
comparatively small disturbances, it still provides intuitive insights on the stability of the power system.
\par 
When subjected to disturbances, the state of the power system deviates from the synchronous state. Hence,
we study the deviation of the frequency and the phase angle difference from the synchronous state, which is the state of the linearized system of the nonlinear power system. We model the disturbances by a Brownian motion process, which is then an input to a linear system, and study the stochastic system 
\begin{subequations}\label{stochasticsystem}
\begin{align}
\text{d}\bm\delta(t)&=\bm \omega(t)\text{d}t,\\
\text{d}\bm \omega(t)&=-\bm M^{-1}\big(\bm L_c\bm \delta(t)+\bm D\bm \omega(t)\big )\text{d}t+\bm M^{-1}\widetilde{\bm B}\text{d}\bm \mu(t),
\end{align}
\end{subequations}
with the state variable, system matrix and input matrix, 
\begin{align}\label{stochasticsystem11}
 \bm x(t)=\begin{bmatrix}
     \bm \delta(t)\\
     \bm\omega(t)
    \end{bmatrix},~~
 \bm A=\begin{bmatrix}
     \bm 0 & \bm I_n\\
    -\bm M^{-1}\bm L_c&-\bm M^{-1}\bm D
   \end{bmatrix},~~
 \bm B &=\begin{bmatrix}
    \bm 0\\
    \bm M^{-1}\widetilde{\bm B}
   \end{bmatrix},
\end{align}
where the notations  $\bm \delta(t),\bm\omega(t)$, $\bm M,\bm D,\bm L_c$ are defined as for (\ref{Linearization}), $\widetilde{\bm B}=\text{diag}(b_i)\in\mathbb{R}^{n\times n}$ where $b_i\in\mathbb{R}$ and $b_i^2$
is used to characterize the strength of the disturbance; $\bm \mu(t)=\text{col}(\mu_i(t))\in\mathbb{R}^n$ is a vector of $n$ independent scalar Brownian motion processes $\mu_i$, which are also all independent of the initial state $\bm x(0)$. A Brownian motion process has increments with a Gaussian probability distribution.  
Here, we refer to $l_{ij}$ as the
\emph{line capacity} of line $e_k$, which is also called the coupling strength between generators, and refer to $l_{c_{ij}}=l_{ij}\cos{\delta_{ij}^*}$ 
as the \emph{weight} of line $e_k$. It is obvious that the weights of the lines are determined by the line capacity and the power flows at the synchronous state.

\par  
In the model (\ref{stochasticsystem}), the disturbances denoted by $\mu_i(t)$ at node $i$ are assumed to be independent, which is reasonable because the locations of the power generators, including renewable power generators, are usually far from each other. Because the system (\ref{stochasticsystem}) is linear, at any time, the probability distribution of the state is Gaussian. To address Problem
\ref{problem1}, we focus on the variance matrices of the frequency and of the phase
angle difference in the invariant probability distribution of the
linear stochastic system, which reflect the dependence of the
fluctuations of the frequency and the phase angle difference on
the system parameters. 
When considering the variance matrix in the invariant probability distribution,
we set the output matrix so that 
\begin{align}\label{output-delta-omega}
\bm y=\bm C\bm x,~~
\bm y=
\begin{bmatrix}
\bm y_\delta\\
\bm y_\omega
\end{bmatrix},
~~
\bm C=
\begin{bmatrix}
\widetilde{\bm C}^{\bm\top}&\bm 0\\
\bm 0&\bm I_n
\end{bmatrix}
\in\mathbb{R}^{(m+n)\times 2n}.
\end{align}
The $m$ elements in $\bm y_\delta$ are the phase angle differences
in the $m$ lines, and the $n$ elements in $\bm y_\omega$ are the frequencies 
at the $n$ nodes. The matrix $\widetilde{\bm C}=(C_{ik})\in\mathbb{R}^{n\times m}$ is the incidence matrix of the network, which
is defined as 
\begin{eqnarray*}
 C_{ik}=\begin{cases}
         +1, &\text{if node $i$ is the begin of line $e_k$},\\
         -1, &\text{if node $i$ is the end of line $e_k$},\\
         0,&\text{otherwise},
        \end{cases}
\end{eqnarray*}
where the direction of line $e_k$ is specified arbitrarily without influence on the study below. 
By the complex network theory \citep{NORMAN}, 
the incidence matrix $\widetilde{\bm C}$ satisfies
\begin{align}\label{Laplacianmatrix}
\widetilde{\bm C}\bm R\widetilde{\bm C}^{\bm\top}=\bm L_c,
\end{align}
where $\bm R=\text{diag}(R_k)\in\mathbb{R}^{m\times m}$ is defined such that 
$R_k=l_{c_{ij}}$ is the weight of line $e_k$ connecting nodes $i$ and $j$.
\par 
Because $\bm x(t)$ is the deviation of the frequency and phase angle difference from the 
synchronous state, it is natural to assume that $\bm x(0)\in G(\bm 0,\bm Q_{x_0})$ where $\bm Q_{x_0}\in\mathbb{R}^{2n\times 2n}$.
Problem 2.2 requires the calculation of the invariant probability
distribution of the deviations of the frequencies and of the phase angle differences, and
requires an analysis of how this distribution depends on the parameters of the power system
in particular on the intensities of the stochastic disturbances. It will be shown in Theorem \ref{lemmacovariance} that the variance matrix in the invariant probability distribution 
is independent of the initial distribution.
Below we restrict attention to the
computation of the invariant probability distribution of the state of the linear stochastic
power system. From that distribution, the variances of the outputs can be computed.
\par 
\section{Derivation of the variance matrices}\label{Section:calculation}
We denote the variance matrix of the frequencies and the phase angle differences at the invariant probability distribution by  
\begin{align*}
\bm Q=
\begin{bmatrix}
\bm Q_\delta& \bm Q_{\delta\omega}^{\bm\top}\\
\bm Q_{\delta\omega}&\bm Q_\omega
\end{bmatrix}\in\mathbb{R}^{(m+n)\times(m+n)},
\end{align*}
where $\bm Q_\delta\in\mathbb{R}^{m\times m}$ denotes the variance matrix of the phase angle differences,  $\bm Q_\omega\in\mathbb{R}^{n\times n}$ denotes the variance matrix of the frequencies, and $\bm Q_{\delta\omega}\in\mathbb{R}^{n\times m}$ denotes the covariance of the phase angle differences and the frequencies. Based on the theory of linear stochastic Gaussian systems, $\bm Q$ is derived by solving a Lyapunov equation, as presented in Definition \ref{appendix1} in the Appendix. However, for a linear stochastic power system, the system 
matrix \emph{$\bm A$ is not Hurwitz}. This is due to the singularity of the Laplacian 
matrix $\bm L_c$, which has a zero eigenvalue. Therefore, the variance matrix $\bm Q$ cannot be 
calculated directly from the corresponding Lyapunov equation. 
A coordinate transformation is required.  Before introducing the transformation, we present a lemma for 
the symmetrizable matrix $\bm M^{-1}\bm L_c$\cite[Appendix]{MLPIAC}. 
\begin{lemma}\label{appendix_theorem}
Consider the Laplacian matrix $\bm L_c$ and the positive-definite diagonal matrix $\bm M^{-1}$ in system (\ref{stochasticsystem}). 
The matrix $\bm L_c$ has a zero eigenvalue with eigenvector $\bm 1_n\in\mathbb{R}^n$ which is a vector with all its elements equal to one and there exists an orthogonal matrix $\bm U\in\mathbb{R}^{n\times n}$ such that
\begin{align}\label{decomposition}
\bm U^{\bm\top}\bm M^{-1/2}\bm L_c\bm M^{-1/2}\bm U=\bm\Lambda_n, 
\end{align}
where $\bm\Lambda_n=\text{diag}(\lambda_i)\in\mathbb{R}^{n\times n}$ with $0= \lambda_1< \lambda_2\cdots\lambda_n$ being the eigenvalues of the matrix $\bm M^{-1/2}\bm L_c\bm M^{-1/2}$, $\bm U=
\begin{bmatrix}
\bm u_1&\bm u_2&\cdots&\bm u_n
\end{bmatrix}$ with $\bm u_i\in\mathbb{R}^n$ being the eigenvector corresponding to $\lambda_i$ for $i=1,\cdots,n$. In addition, $\bm u_1=\sigma\bm M^{1/2}\bm 1_n$ where $\sigma$ is a constant. 
\end{lemma}
\par 
Based on Lemma \ref{appendix_theorem}, we transform the coordinates
of $(\bm \delta,\bm \omega)$ into the eigen-space as follows. 
Let $\bm x_1=(\bm M^{-1/2}\bm U)^{-1}\bm \delta$, $\bm x_2=(\bm M^{-1/2}\bm U)^{-1}\bm \omega$ and insert (\ref{decomposition}) into
(\ref{stochasticsystem}), we derive
\begin{subequations}\label{transformed}
\begin{align}
\hspace{-10pt}
\text{d}\bm x_1&=\bm x_2\text{d}t,\\
\hspace{-10pt}
\text{d}\bm x_2&=-\big(\bm \Lambda_n \bm x_1+\bm U^{\bm\top}\bm M^{-1}\bm D\bm U\bm x_2\big)\text{d}t+\bm U^{\bm\top}\bm M^{-1/2}\widetilde{\bm B}\text{d}{\bm \mu(t)},
\end{align}
\end{subequations}
with the state variable, system matrix and input matrix becoming
\begin{align}
 \bm x_e=\begin{bmatrix}
     \bm x_1\\
     \bm x_2
    \end{bmatrix},~
 \bm A_e=\begin{bmatrix}
     \bm 0 & \bm I_n\\
    -\bm \Lambda_n &-\bm U^{\bm\top}\bm M^{-1}\bm D\bm U
   \end{bmatrix},  
\bm B_e =\begin{bmatrix}
    \bm 0\\
    \bm U^{\bm\top}\bm M^{-1/2}\widetilde{\bm B}
   \end{bmatrix},
    \label{blockmatrix0}
\end{align}
and initial distribution $\bm x_e(0)\in G(\bm 0,\bm Q_{x_{e_0}})$ such that
\begin{align*}
\bm Q_{x_{e_0}}&=\bm T
   \bm Q_{x_0}
   \bm T^{\bm \top}\in\mathbb{R}^{2n\times 2n},\\
\bm T&=
\begin{bmatrix}
     (\bm M^{-1/2}\bm U)^{-1} & \bm 0\\
    \bm 0 &(\bm M^{-1/2}\bm U)^{-1}
   \end{bmatrix}\in\mathbb{R}^{2n\times 2n}.
\end{align*}
The output (\ref{output-delta-omega})
becomes
\begin{align}\label{transformed-delta-omega}
\bm y=\bm C_e\bm x_e, ~~\bm C_e=
\begin{bmatrix}
\widetilde{\bm C}^{\bm\top} \bm M^{-1/2}\bm U&\bm 0\\
\bm 0&\bm M^{-1/2}\bm U
\end{bmatrix}\in\mathbb{R}^{(m+n)\times 2n}.
\end{align}
Because $\widetilde{\bm C}$ is an incidence matrix of the network, it satisfies 
$\widetilde{\bm C}^{\bm\top} \bm 1_n=\bm 0$. Thus, $\widetilde{\bm C}^{\bm\top} \bm M^{-1/2}\bm u_1=\bm 0$ since $\bm u_1 =\sigma\bm M^{1/2} \bm 1_n$, which leads to 
\begin{align*}
\widetilde{\bm C}^{\bm\top}\bm M^{-1/2}\bm U=
\begin{bmatrix}
\bm 0&\widetilde{\bm C}^{\bm\top}\bm M^{-1/2}\bm u_2&\cdots&\widetilde{\bm C}^{\bm\top}\bm M^{-1/2}\bm u_n
\end{bmatrix},
\end{align*}
where the entries in the first column are all zero. So the entries in 
the first column of $\bm C_e$ are all zero. 
Because the diagonal matrix $\bm \Lambda_n$ has a zero entry at position $(1,1)$,  the entries of the first column of $\bm A_e$ are also all zero.  In addition, the entries of the first row of $\bm B_e$ are all zero. Hence, we decompose the system matrix $\bm A_e$, the input matrix $\bm B_e$, and the output matrix $\bm C_e$ into
\begin{align}\label{blockmatrix}
 \bm A_e=\begin{bmatrix}
         0 & \bm A_{12}\\
     \bm 0 &\bm A_2
   \end{bmatrix}, ~~
  \bm B_e =\begin{bmatrix}
    \bm 0\\
    \bm B_2
   \end{bmatrix}, ~~
   \bm C_e=
   \begin{bmatrix}
   \bm 0&\bm C_2
   \end{bmatrix},
\end{align}
where $\bm A_{12}\in \mathbb{R}^{1\times(2n-1)}$ and $\bm A_2\in \mathbb{R}^{(2n-1)\times (2n-1)}$, $\bm B_2\in\mathbb{R}^{(2n-1)\times n}$ and $\bm C_2$ is the matrix obtained by removing the 
first column of the output matrix in (\ref{transformed-delta-omega}) so that 
\begin{align}\label{C2-delta-omega}
\bm C_2=
\begin{bmatrix}
\widetilde{\bm C}^{\bm\top}\bm M^{-1/2}\widehat{\bm U}&\bm 0\\
\bm 0 &\bm M^{-1/2}\bm U
\end{bmatrix}
\in\mathbb{R}^{(m+n)\times (2n-1)},
\end{align}
with $\widehat{\bm U}=
\begin{bmatrix}
\bm u_2&\bm u_3&\cdots&\bm u_n
\end{bmatrix}
\in\mathbb{R}^{n\times (n-1)}$. According to these decompositions, the matrix $\bm Q_{x_{e_0}}$ 
is further rewritten as 
\begin{align}\label{initialQe}
\bm Q_{x_{e_0}}=
\begin{bmatrix}
Q_{e_1}&\bm Q_{e_{12}}^{\bm \top}\\
\bm Q_{e_{12}}&\bm Q_{e_2}
\end{bmatrix},
\end{align}
where $Q_{e_1}\in\mathbb{R},~\bm Q_{e_{12}}\in\mathbb{R}^{2n-1},~\bm Q_{e_2}\in\mathbb{R}^{(2n-1)\times (2n-1)}$.
\par 
In (\ref{blockmatrix}), $\bm A_2$ is obtained from $\bm A_e$ by removing the first column and the first row
and $\bm B_2$ is obtained from $\bm B_e$ by removing the first row.  
Since the eigenvalues of $\bm A_e$ all have non-positive real parts and $\text{rank}(\bm A_e)=2n-1$,  $\bm A_2$ is Hurwitz. 
With (\ref{blockmatrix0}) and (\ref{blockmatrix}), $\bm A_2,$ and $\bm B_2$ are further written into block matrices,
\begin{align}\label{matrix-A2}
\bm A_2=
\begin{bmatrix}
\bm 0 &\bm A_{22}\\
\bm A_{23} &\bm A_{24}
\end{bmatrix},~~
\bm B_2=
\begin{bmatrix}
\bm 0\\
\bm B_{22}
\end{bmatrix},
\end{align}
where
\begin{subequations}\label{matrix-A22}
\begin{align}
&\bm A_{22}=
\begin{bmatrix}
\bm 0&\bm I_{n-1}
\end{bmatrix}
\in\mathbb{R}^{(n-1)\times n}, ~~
\bm A_{23}^{\bm\top}=
\begin{bmatrix}
\bm 0&-\bm \Lambda_{n-1}
\end{bmatrix}
\in\mathbb{R}^{(n-1)\times n},\\
&\bm A_{24}=-\bm U^{\bm\top}\bm M^{-1}\bm D\bm U\in\mathbb{R}^{n\times n},~~\bm B_{22}=\bm U^{\bm\top}\bm M^{-1/2} \widetilde{\bm B}\in\mathbb{R}^{n\times n}.
\end{align}
\end{subequations}
Here, $\bm \Lambda_{n-1}=\text{diag}(\lambda_i,i=2,\cdots,n)\in\mathbb{R}^{(n-1)\times(n-1)}$ is obtained by removing the first column and the first row of the diagonal matrix $\bm \Lambda_n$. 
With the above notations, for the variance matrix of the output of the system (\ref{stochasticsystem}), we have 
the following theorem.
\begin{theorem}\label{lemmacovariance}
The variance matrix $\bm Q$ of the output $\bm y$ of the system (\ref{stochasticsystem}) in the invariant probability distribution  satisfies
\begin{align}\label{Qy}
\bm Q=\bm C_2\bm Q_x\bm C_2^{\bm\top},
\end{align}
where $\bm C_2$ is defined in (\ref{C2-delta-omega}), $\bm Q_x\in\mathbb{R}^{(2n-1)\times (2n-1)}$ is the unique solution of the following Lyapunov equation
\begin{align}\label{barQ}
\bm A_2\bm Q_x+\bm Q_x\bm A_2^{\bm\top}+\bm B_2\bm B_2^{\bm\top}=\bm 0,
\end{align}
where $\bm A_2,~~\bm B_2$ are defined in (\ref{matrix-A2}) with blocks in (\ref{matrix-A22}). 
\end{theorem}
\emph{Proof:} We decompose the state variable $\bm x_e=(x_{e_1},\bm x_{e_2}^{\bm\top})^{\bm\top}$ with 
$x_{e_1}\in\mathbb{R}$, $\bm x_{e_2}\in\mathbb{R}^{2n-1}$. From (\ref{transformed}-\ref{blockmatrix0}) and 
the decomposition of matrices in (\ref{blockmatrix}), we obtain the stochastic process 
\begin{align}\label{Qe}
\text{d}\bm x_{e_2}(t)=\bm A_2 \bm x_{e_2}(t)\text{d}t+\bm B_2\text{d}\bm \mu(t),
\end{align}
where $\bm A_2$ is Hurwitz. 
From (\ref{blockmatrix}), it is seen that the entries in the first column of $\bm C_e$ are all zero. Thus, the output $\bm y(t)$ satisfies
\begin{align}\label{Qey}
\bm y(t)=\bm C_e\bm x_e(t)=\bm C_2\bm x_{e_2}(t),
\end{align}
From (\ref{initialQe}), we obtain the initial value of $\bm x_{e_2}(0)\in G(\bm 0,\bm Q_{e_2})$.  
Consider the stochastic process (\ref{Qe}) with output in (\ref{Qey}). Following Definition \ref{appendix1} in the Appendix,
the variance of the output $\bm y(t)$ is 
\begin{align*}
\bm Q_y(t)=\bm C_2\text{e}^{\bm A_2 t}\bm Q_{e_2}\text{e}^{\bm A_2^{\bm \top} t}\bm C_2^{\bm \top}+\int_0^{t}{\bm C_2\text{e}^{\bm A_2 \tau}\bm B_2\bm B_2^{\bm \top}\text{e}^{\bm A_2^{\bm\top} \tau}\bm C_2^{\bm \top}}\text{d}\tau. 
\end{align*}
With the Hurwitz condition of $\bm A_2$, we obtain the variance matrix of $\bm y(t)$,  
\begin{align*}
\bm Q=\lim_{t\rightarrow}\bm Q_y(t)=\int_0^{+\infty}{\bm C_2\text{e}^{\bm A_2 \tau}\bm B_2\bm B_2^{\bm \top}\text{e}^{\bm A_2^{\bm\top} \tau}\bm C_2^{\bm \top}}\text{d}\tau,
\end{align*}
which can be solved from (\ref{Qy}) with
\begin{align*}
\bm Q_x=\int_0^{+\infty}{\text{e}^{\bm A_2 \tau}\bm B_2\bm B_2^{\bm \top}\text{e}^{\bm A_2^{\bm\top} \tau}}\text{d}\tau,
\end{align*}
which is the Controllability Gramian of the pair $(\bm A_2,\bm B_2)$ and is the unique solution 
of the Lyapunov equation (\ref{barQ}). 
\hfill $\square$
\par 
It is seen that the invariant matrix $\bm Q$ is independent of the initial distribution of the original process $\bm x(t)$
defined in (\ref{stochasticsystem}-\ref{stochasticsystem11}). With Theorem \ref{lemmacovariance} and the formulation 
of $\bm A_2$ and $\bm B_2$ in (\ref{matrix-A2}-\ref{matrix-A22}), the variance matrix $\bm Q_x$ can be obtained by solving the Lyapunov equation using Matlab and the variance matrix $\bm Q$ can be further calculated from (\ref{Qy}).
Clearly, the larger the variances, the more serious the fluctuations in the nodes 
and lines will be. Thus, from the diagonal elements of $\bm Q$, the vulnerable nodes and lines with large variances 
can be identified.
\begin{remark}
If $\bm Q_\delta$ is needed only,
the output is set  for the system (\ref{transformed}) as 
\begin{align*}
\bm y=\bm C_e\bm x_e, ~~\bm C_e=
\begin{bmatrix}
\widetilde{\bm C}^{\bm\top}\bm M^{-1/2}\bm U&\bm 0
\end{bmatrix}
\in\mathbb{R}^{m\times 2n}. 
\end{align*}
By removing the first column of $\bm C_e$,  we obtain 
\begin{align}\label{C2-delta}
\bm C_2=
\begin{bmatrix}
\widetilde{\bm C}^{\bm\top}\bm M^{-1/2}\widehat{\bm U}&\bm 0
\end{bmatrix}
\in\mathbb{R}^{m\times (2n-1)}
\end{align}
for the calculation of $\bm Q_\delta$
by (\ref{Qy}). 
\par 
If $\bm Q_\omega$ is needed only,
the output is set for the system (\ref{transformed}) as 
\begin{align*}
\bm y=\bm C_e\bm x_e,  ~~\bm C_e=
\begin{bmatrix}
\bm 0&\bm M^{-1/2}\bm U
\end{bmatrix}
\in\mathbb{R}^{n\times 2n}. 
\end{align*}
By removing the first column of $\bm C_e$, we obtain
\begin{align}\label{C2-omega}
\bm C_2=
\begin{bmatrix}
\bm 0&\bm M^{-1/2}\bm U
\end{bmatrix}
\in\mathbb{R}^{n\times (2n-1)}
\end{align}
for the calculation of $\bm Q_\omega$ by (\ref{Qy}). 
\end{remark}
\par 
The variance of the frequency at a node can also be calculated 
via the $\mathcal{H}_2$ norm of the input-output system with the output being 
the frequency at this node. However, when considering the variances of the frequencies at all the nodes, 
$n$ Lyapunov equations need to be solved. Similarly, when considering the variances 
of the phase angle differences, the solution of $m$ Lyapunov equations are required. These computations have a high computational complexity. Furthermore, the correlation of the outputs cannot be derived in this way.

\section{Explicit formulas of the variance matrices for networks with uniform disturbance-damping ratio}\label{Section:formula}

Based on the following assumption, we derive the explicit formula
of the solution $\bm Q$. 
\begin{assumption}\label{assumption}
Consider the stochastic system (\ref{stochasticsystem}).
Assume that the uniform disturbance-damping ratio holds, in which there exists 
a strictly positive number $\eta\in(0,+\infty)$ such that for all nodes $i\in\mathcal{V}$, $b_i^2/d_i=\eta$.  
\end{assumption}
In practice, in order to achieve fair power sharing, the drooping coefficients $d_i$ are often scheduled proportionally to the rating of the power source. Thus, it is reasonable to expect that the strength of the disturbance, that is characterized by $b_i^2$, is proportional to the rating of the power source.
On contrast to this assumption, one says that the non-uniform disturbance-damping ratio holds in the
complementary case, or, equivalently, if there exist $i,~j\in\mathcal{V}$ with $i\neq j$ such that $b_i^2/d_i\neq b_j^2/d_j$.
\par 
To compute the variance matrix $\bm Q$ one has to first compute the variance matrix $\bm Q_x$ as stated next. 
\begin{lemma}\label{lemma1}
We write the matrix  $\bm Q_x$ defined in Theorem \ref{lemmacovariance} into a block matrix, 
\begin{align*}
\bm Q_x=
\begin{bmatrix}
\bm Q_{1}&\bm Q_{2}\\
\bm Q_{2}^{\bm\top}&\bm Q_{3}
\end{bmatrix},
\end{align*}
where $\bm Q_{1}\in\mathbb{R}^{(n-1)\times(n-1)}$, $\bm Q_{2}\in\mathbb{R}^{(n-1)\times n}$ and $\bm Q_{3}\in\mathbb{R}^{n\times n}$. If Assumption \ref{assumption} holds and $\bm Q_x$ satisfies the Lyapunov equation (\ref{barQ}), then 
\begin{align}\label{solution}
\bm Q_{1}=\frac{1}{2}\eta\bm \Lambda_{n-1}^{-1},~~
\bm Q_{2}=\bm 0,~~
\bm Q_{3}&=\frac{1}{2}\eta\bm I_n,
\end{align}
where $\bm \Lambda_{n-1}$ is obtained from 
the matrix $\bm \Lambda_n$ by removing the first column and the first row as in (\ref{matrix-A22}).
\end{lemma}
\emph{Proof:}
With the block matrices $\bm A_2$ and $\bm B_2$ in (\ref{matrix-A2}) and the corresponding blocks $\bm A_{22}$, $\bm A_{23}$, $\bm A_{24}$ and $\bm B_{22}$ in (\ref{matrix-A22}), we derive 
from the Lyapunov equation (\ref{barQ}) that 
\begin{align*}
\begin{bmatrix}
\bm 0&\bm A_{22}\\
\bm A_{23}&\bm A_{24}
\end{bmatrix}
\begin{bmatrix}
\bm Q_{1}&\bm Q_{2}\\
\bm Q_{2}^{\bm\top}&\bm Q_{3}
\end{bmatrix}
&+
\begin{bmatrix}
\bm Q_{1}&\bm Q_{2}\\
\bm Q_{2}^{\bm\top}&\bm Q_{3}
\end{bmatrix}
\begin{bmatrix}
\bm 0&\bm A_{22}\\
\bm A_{23}&\bm A_{24}
\end{bmatrix}^{\bm\top}\\
&+
\begin{bmatrix}
\bm 0\\
\bm B_{22}
\end{bmatrix}
\begin{bmatrix}
\bm 0&\bm B_{22}^{\bm\top}
\end{bmatrix}
=\bm 0
\end{align*}
which yields
\begin{subequations}\label{equivalenteq}
\begin{align}
\bm Q_{2}\bm A_{22}^{\bm\top}+\bm A_{22}\bm Q_{2}^{\bm\top}&=\bm 0,\\
\bm Q_{1}\bm A_{23}^{\bm\top}+\bm Q_{2}\bm A_{24}^{\bm\top}+\bm A_{22}\bm Q_{3}&=\bm 0,\label{equivalent2}\\
\bm Q_{2}^{\bm\top}\bm A_{23}^{\bm\top}+\bm Q_{3}\bm A_{24}^{\bm\top}+\bm A_{23}\bm Q_{2}+\bm A_{24}\bm Q_{3}&=-\bm B_{22}\bm B_{22}^{\bm\top}.\label{equivalent3}
\end{align}
\end{subequations}
The idea to solve the above equations is as follows. 
We first assume $\bm Q_{2}=\bm 0$, then solve $\bm Q_{3}$ and $\bm Q_{1}$ from (\ref{equivalent3}) and (\ref{equivalent2}) respectively, finally we check whether these three matrices satisfy all the equations in (\ref{equivalenteq}). If that is true, then from the uniqueness of the solution of (\ref{barQ}), we have obtained the solution $\bm Q_x$
for (\ref{barQ}).  From (\ref{equivalent3}) with the 
formula for $\bm A_{24}$ and $\bm B_{22}$ in (\ref{matrix-A22}) and $\bm Q_2=\bm 0$, we derive 
\begin{align*}
\bm Q_{3}\bm U^{\bm\top}\bm M^{-1}\bm D\bm U+\bm U^{\bm\top}\bm M^{-1}\bm D\bm U\bm Q_{3}=
\bm U^{\bm\top}\bm M^{-1/2} \widetilde{\bm B}\widetilde{\bm B}^{\bm\top} \bm M^{-1/2}\bm U
\end{align*} 
which has a unique solution
\begin{align*}
\bm Q_{3}=\frac{1}{2}\bm U^{\bm\top}\bm D^{-1}\widetilde{\bm B}^{2}\bm U,
\end{align*}
where the fact that $\bm M,~~\bm D,~~\widetilde{\bm B}$ are diagonal matrices and $\widetilde{\bm B}^2=\widetilde{\bm B}\widetilde{\bm B}^{\bm\top}$ are used.  It is obvious that the diagonal entries 
of $\bm D^{-1}\widetilde{\bm B}^2$ are $b_i^{2}/d_i=\eta$ for $i=1,\cdots,n$, which yields $\bm D^{-1}\widetilde{\bm B}^2=\eta \bm I_n$. Thus $\bm Q_3=\frac{1}{2}\eta\bm I_n$. From (\ref{equivalent2}) with the formulas for $\bm A_{23}$ and $\bm A_{22}$ in (\ref{matrix-A22}) and $\bm Q_2=\bm 0$, 
we derive 
\begin{align*}
\bm Q_{1}
\begin{bmatrix}
\bm 0&-\bm \Lambda_{n-1}
\end{bmatrix}
+
\frac{1}{2}\eta\begin{bmatrix}
\bm 0&\bm I_{n-1}
\end{bmatrix}
 \bm I_n
=\bm 0,
\end{align*}
which leads to
\begin{align*}
-\bm Q_1\bm\Lambda_{n-1}+\frac{1}{2}\eta \bm I_{n-1}=\bm 0 .
\end{align*}
Thus, $\bm Q_1=\frac{1}{2}\eta\bm \Lambda^{-1}_{n-1}$. In conclusion, by assuming $\bm Q_2=\bm 0$, we have obtained the 
explicit formulas for $\bm Q_1$ and $\bm Q_3$ as presented in (\ref{solution}). Furthermore, it can be verified 
that $\bm Q_1,\bm Q_2$ and $\bm Q_3$ satisfy (\ref{equivalenteq}) which is equivalent 
to the Lyapunov equation (\ref{barQ}). \hfill $\square$
\par 
By Lemma \ref{lemma1}, we derive the independence
of the stochastic process of the frequency to the phase angle differences in the lines. 
In addition, 
an explicit formula for the variance matrix $\bm Q_\omega$ of the frequencies at the nodes is deduced. 
\par 
\begin{theorem}\label{theorem-omega}
Consider the system (\ref{stochasticsystem}) with Assumption \ref{assumption}. 
\begin{enumerate}[(i)]
\item The invariant probability distributions of the frequencies and of the phase angle differences are independent, i.e., $\bm Q_{\delta\omega}=\bm 0$. 
\item The variance matrix 
of the frequencies is 
\begin{align}\label{theorem0-1}
\bm Q_{\omega}=\frac{1}{2}\eta\bm M^{-1}.
\end{align}
\end{enumerate}
\end{theorem}
\emph{Proof:} 
(i)  We take $\bm C_2$ in (\ref{C2-delta-omega}) as the output matrix for the system (\ref{transformed}).  By 
Theorem \ref{lemmacovariance},  we obtain that the variance 
matrix $\bm Q$ satisfies\begin{align*}
\bm Q&=\bm C_2\bm Q_x\bm C_2^{\bm\top}\\
&=
\begin{bmatrix}
 \frac{\eta}{2}\widetilde{\bm C}^{\bm\top}\bm M^{-1/2}\widehat{\bm U}\bm \Lambda_{n-1}^{-1}\widehat{\bm U}^{\bm\top}\bm M^{-1/2}\widetilde{\bm C}&\bm 0\\
\bm 0  &\frac{\eta}{2}\bm M^{-1/2}\bm U\bm U^{\bm\top}\bm M^{-1/2}
\end{bmatrix}
\end{align*}
from which we obtain that the blocks of $\bm Q$ satisfy
\begin{align}
&\bm Q_\delta=\frac{\eta}{2}\widetilde{\bm C}^{\bm\top}\bm M^{-1/2}\widehat{\bm U}\bm \Lambda_{n-1}^{-1}\widehat{\bm U}^{\bm\top}\bm M^{-1/2}\widetilde{\bm C},\label{Q-delta}\\
&\bm Q_\omega=\frac{\eta}{2}\bm M^{-1/2}\bm U\bm U^{\bm\top}\bm M^{-1/2},\label{Q-omega}\\
\nonumber
&\bm Q_{\delta\omega}=\bm 0.
\end{align}
Since the off-diagonal block matrix $\bm Q_{\delta\omega}$
of the variance matrix $\bm Q$ of the output (\ref{output-delta-omega}) is a zero matrix and the stochastic process (\ref{stochasticsystem11}) is Gaussian, the invariant
probability distribution of the frequency and the phase angle differences are 
independent. 
\par 
(ii) Given the fact that $\bm U$ is an orthogonal 
matrix and $\bm M$ is a diagonal matrix, (\ref{Q-omega}) is simply rewritten into (\ref{theorem0-1}).
  \hfill $\square$
\par 
In the proof, the fact is applied that two Gaussian distributed random variables are independent if and only if 
their covariance equals to zero. 
Due to the independence between the frequencies and the phase angle differences, Theorem \ref{theorem-omega}(i) indicates that the fluctuations of the frequencies and those of the phase-angle differences have no stochastic
relation with each other.

Formula (\ref{theorem0-1})  is verified in an example presented in Section \ref{casestudy}.
This formula shows the dependence of the variances of the frequencies at the nodes on the system parameters.  First, the variance matrix $\bm Q_\omega$ is a diagonal matrix with $\bm M=\text{diag}(m_i)\in\mathbb{R}^{n\times n}$; thus, the frequencies in different nodes are independent in the invariant probability distribution.  Second, the variance of the frequency at each node increases linearly with the disturbance-damping ratio and is inversely proportional to the inertia of the synchronous machine at this node. This shows the importance of the inertia and the damping coefficient in suppressing the frequency deviation in the power network. However, increasing the inertia at a node suppresses the fluctuations of the frequency only at this node, without any effect on the other nodes. The vulnerable nodes are the ones with small inertia. Those nodes
will then have large variances and these variances are not influenced by the disturbances at
the other nodes. Finally, the parameters, the power generation and the loads, which determine the synchronous state $(\bm \delta^*,\bm 0)$ and play roles in determining the value $l_{c_{ij}}$ as shown in (\ref{Laplacianform}), the line capacity and the network topology are all absent from the formula. It is surprising that these parameters have no impact on the variances of the frequencies.  This might be 
due to the assumption of uniform disturbance-damping ratio in Assumption \ref{assumption}. 
Whether this occurs in the systems without the assumption still needs further study. 
\par 
The trace of $\bm Q_\omega$ is derived directly from (\ref{theorem0-1}) as presented in the following corollary, which is actually the $\mathcal{H}_2$ norm of a linear input-output system \citep{optimal_inertia_placement}.
\begin{corollary}
From Theorem \ref{theorem-omega} one obtains that 
\begin{align*}
\text{trace}(\bm Q_\omega)= \frac{\eta}{2}\sum_{i=1}^{n}\frac{1}{m_i}. 
\end{align*}
\end{corollary}
In order to reveal the dependence of the variances of the phase angle differences on the system parameters, we further deduce the formula of $\bm Q_\delta$ based on Lemma \ref{lemma1}.
Before presenting this explicit formula, we first introduce a lemma for the properties of the matrix $\widehat{\bm C}=\bm R^{1/2}\widetilde{\bm C}^{\bm\top}\bm M^{-1/2}$.
\begin{lemma}\label{lemma2}
Consider the symmetric matrix $\widehat{\bm C}\widehat{\bm C}^{\bm\top}$ with
 $\widehat{\bm C}=\bm R^{1/2}\widetilde{\bm C}^{\bm\top}\bm M^{-1/2}$. There exists an orthogonal matrix $\bm W\in\mathbb{R}^{m\times m}$
 for $m\geq n$
 such that 
\begin{align}\label{lemma2-1}
\bm W^{\bm\top}\widehat{\bm C}\widehat{\bm C}^{\bm\top}\bm W=\bm \Lambda_m,~~
\bm \Lambda_m=
\begin{bmatrix}
\bm\Lambda_{n-1}&\bm 0\\
\bm 0&\bm 0 
\end{bmatrix}\in\mathbb{R}^{m\times m}
\end{align}
with $\bm \Lambda_{n-1}$ defined in Lemma {\ref{lemma1}}. If we denote $\bm W=[\bm w_1,\bm w_2,\cdots,\bm w_{n-1},\bm w_n,\cdots,\bm w_m]$,
then the vector $\bm w_i$ is the orthonormal
eigenvector of $\widehat{\bm C}\widehat{\bm C}^{\bm\top}$ corresponding to 
the nonzero eigenvalue $\lambda_{i+1}$ for $i=1,\cdots,n-1$  and $\bm w_i$ for $i=n,\cdots,m$ are
the orthonormal eigenvectors corresponding to the zero eigenvalue. For the case with $m=n-1$, all the eigenvalues of $\widehat{\bm C}\widehat{\bm C}^{\bm\top}$ are non-zero, and
\begin{align*}
\bm \Lambda_m=\bm \Lambda_{n-1},~~~~\bm W=[\bm w_1,\bm w_2,\cdots,\bm w_{n-1}].
\end{align*} 
\end{lemma}
\emph{Proof:} 
For a connected graph, we have $\text{rank}(\widetilde{\bm C})=n-1$, 
which leads to $\text{rank}(\widehat{\bm C})=n-1$. 
Since the kernel of $\widehat{\bm C}\widehat{\bm C}^{\bm\top}\bm X=\bm 0$
and $\widehat{\bm C}^{\bm\top}\bm X=0$ are identical, $\text{rank}(\widehat{\bm C}\widehat{\bm C}^{\bm\top})=n-1$.  Based on Theorem \ref{Lemma0},  we only need to prove that
the non-zero
diagonal elements of $\bm \Lambda_m$ are the non-zero eigenvalues of $\widehat{\bm C}\widehat{\bm C}^{\bm\top}$.
We obtain from (\ref{Laplacianmatrix},\ref{decomposition}) that 
\begin{align*}
\bm U^{\bm\top}\widehat{\bm C}^{\bm\top}\widehat{\bm C}\bm U
&=\bm U^{\bm\top}\bm M^{-1/2}\widetilde{\bm C}\bm R\widetilde{\bm C}^{\bm\top} M^{-1/2}\bm U\\
&=\bm U^{\bm\top}\bm M^{-1/2}\bm L_c \bm M^{-1/2}\bm U
=\bm\Lambda_n.
\end{align*}
With the left multiplication of $\widehat{\bm C}\bm U$ to the above equation, we obtain 
\begin{align}\label{eigenequation}
\widehat{\bm C}\widehat{\bm C}^{\bm\top}\widehat{\bm C}\bm U=\widehat{\bm C}\bm U\bm \Lambda_n.
\end{align}
We write $\bm U$ into the form 
$\begin{bmatrix}
\bm u_1&\bm u_2&\bm u_3&\cdots&\bm u_n
\end{bmatrix}$. 
From Lemma \ref{appendix_theorem} and $\widetilde{\bm C}^{\bm\top} \bm 1_n=\bm 0$, we obtain  $\widetilde{\bm C}^{\bm\top}\bm M^{-1/2}\bm u_1=\bm 0$, which leads to $\widehat{\bm C}\bm u_1=\bm 0$. 
Hence, we derive from (\ref{eigenequation}) that
\begin{align*}
\widehat{\bm C}\widehat{\bm C}^{\bm\top}
\begin{bmatrix}
\widehat{\bm C}\bm u_2,\widehat{
\bm C}\bm u_3,\cdots,\widehat{\bm C}\bm u_n
\end{bmatrix}
=
\begin{bmatrix}
\lambda_2\widehat{\bm C}\bm u_2&\lambda_3\widehat{\bm C}\bm u_2&\cdots&\lambda_n\widehat{\bm C}\bm u_n
\end{bmatrix},
\end{align*}
which indicates that $\lambda_i$ and  $\widehat{\bm C}\bm u_i$ for $i=2,\cdots,n$ are 
the eigenvalues and the corresponding eigenvectors of the matrix $\widehat{\bm C}\widehat{\bm C}^{\bm\top}$. 
 \hfill $\square$
\par 
Based on Lemma \ref{lemma2}, we present the explicit formula for the 
variance matrix $\bm Q_\delta$ in the following theorem. 
\begin{theorem}\label{theoremmain}
Consider the system (\ref{stochasticsystem}) with Assumption \ref{assumption}. 
The variance matrix of the phase angle differences in the invariant probability distribution satisfies
\begin{align}\label{theoremmain-1}
\bm Q_\delta=\frac{1}{2}\eta\bm R^{-1/2}\big(\bm I_m-\sum_{i=1}^{m-n+1}\bm X_i\bm X_i^{\bm\top}\big)\bm R^{-1/2}
\end{align} 
where $\{\bm X_i\in\mathbb{R}^{m},i=1,2,\cdots,m-n+1\}$ is an orthonormal basis vector of the kernel of the matrix $\widetilde{\bm C}\bm R^{1/2}$ such that $\widetilde{\bm C}\bm R^{1/2}\bm X_i=\bm 0$.  
Clearly, because the inertia values are absent 
from the formula, they have no impact on the variance of the phase angle difference in each line.
\end{theorem}
\emph{Proof:} From (\ref{Q-delta}), we obtain
\begin{align}
\nonumber
\bm Q_\delta&=\frac{\eta}{2}\widetilde{\bm C}^{\bm\top}\bm M^{-1/2}\widehat{\bm U}\bm \Lambda_{n-1}^{-1}\widehat{\bm U}^{\bm\top}\bm M^{-1/2}\widetilde{\bm C}\\
\nonumber 
&~~\text{\small by Theorem \ref{Lemma0}}\\
\nonumber
&=\frac{\eta}{2}\widetilde{\bm C}^{\bm\top}\bm M^{-1/2}\big(\bm M^{-1/2}\bm L_c \bm M^{-1/2}\big)^{\dag}\widehat{\bm U}\widehat{\bm U}^{\bm\top}\bm M^{-1/2}\widetilde{\bm C}\\
\nonumber
&~~\text{\small by $[\bm u_1,\widehat{\bm U}][\bm u_1,\widehat{\bm U}]^\top=\bm u_1\bm u_1^\top+\widehat{\bm U}\widehat{\bm U}^\top=\bm I_n$}\\
\nonumber
&=\frac{\eta}{2}\widetilde{\bm C}^{\bm\top}\bm M^{-1/2}\big(\bm M^{-1/2}\bm L_c \bm M^{-1/2}\big)^{\dag}(\bm I_n-\bm u_1\bm u_1^{\bm\top})\bm M^{-1/2}\widetilde{\bm C}\\
\nonumber
&~~\text{\small by $\widetilde{\bm C}^\top \bm M^{-1/2}\bm u_1=\bm 0$ obtained from Lemma \ref{appendix_theorem}}\\
&=\frac{\eta}{2}\widetilde{\bm C}^{\bm\top}\bm M^{-1/2}\big(\bm M^{-1/2}\bm L_c \bm M^{-1/2}\big)^{\dag}\bm M^{-1/2}\widetilde{\bm C}\label{proof-theoremmain-0}\\
\nonumber 
&~~\text{\small by  (\ref{Laplacianmatrix})}\\
\nonumber
&=\frac{\eta}{2}\widetilde{\bm C}^{\bm\top}\bm M^{-1/2}\big(\bm M^{-1/2} \widetilde{\bm C}\bm R\widetilde{\bm C}^{\bm\top}\bm M^{-1/2}\big)^{\dag}\bm M^{-1/2}\widetilde{\bm C}
\end{align}
where $(\cdot)^{\dag}$ denotes the Moore-Penrose pseudo inverse of a matrix. 
With $\widehat{\bm C}=\bm R^{1/2}\widetilde{\bm C}^{\bm\top}\bm M^{-1/2}$ as in Lemma \ref{lemma2}, we further obtain 
\begin{align}\label{proof-theoremmain-1}
\bm Q_\delta=\frac{\eta}{2}\bm R^{-1/2}\widehat{\bm C}\Big(\widehat{\bm C}^{\bm\top}\widehat{\bm C}\Big)^{\dag}\widehat{\bm C}^{\bm\top}\bm R^{-1/2}.
\end{align}
By Lemma \ref{lemma2} and left multiplying (\ref{lemma2-1}) by $\widehat{\bm C}^{{\bm\top}}\bm W$, we get
\begin{align*}
\widehat{\bm C}^{\bm\top}\widehat{\bm C}\widehat{\bm C}^{{\bm\top}}\bm W=\widehat{\bm C}^{{\bm\top}}\bm W\bm\Lambda_m,
\end{align*}
which indicates that the column vectors of $\widehat{\bm C}^{{\bm\top}}\bm W$ are the eigenvectors 
of $\widehat{\bm C}^{\bm\top}\widehat{\bm C}$. We focus on the first $n-1$ eigenvectors $\widehat{\bm C}^{\bm\top}\bm w_1,\cdots, \widehat{\bm C}^{\bm\top}\bm w_{n-1}$ 
in matrix $\widehat{\bm C}^{{\bm\top}}\bm W$, which are orthogonal.
The normalization of $\widehat{\bm C}^{\bm\top}\bm w_i$ for $i=1,\cdots, n-1$ 
yields
\begin{align*}
\lambda_{2}^{-1/2}\widehat{\bm C}^{\bm\top}\bm w_1,\lambda_{3}^{-1/2}\widehat{\bm C}^{\bm\top}\bm w_2,\cdots, \lambda_{n}^{-1/2}\widehat{\bm C}^{\bm\top}\bm w_{n-1}.
\end{align*}
With these unit vectors, we obtain  from Theorem \ref{Lemma0} that the Moore-Penrose pseudo inverse of $ \widehat{\bm C}^{{\bm\top}}\widehat{\bm C}$ satisfies
\begin{align*}
\Big( \widehat{\bm C}^{{\bm\top}}\widehat{\bm C}\Big)^{\dag}
=\sum_{i=2}^{n}{\frac{1}{\lambda_i^2}(\widehat{\bm C}^{\bm\top}\bm w_{i-1})(\widehat{\bm C}^{\bm\top}\bm w_{i-1})^{{\bm\top}}}.
\end{align*}
With (\ref{lemma2-1}), we further obtain 
\begin{align*}
\widehat{\bm C}\Big(\widehat{\bm C}^{{\bm\top}}\widehat{\bm C}\Big)^{\dag}\widehat{\bm C}^{{\bm\top}}&=
\sum_{i=2}^{n}{\frac{1}{\lambda_{i}^2}\widehat{\bm C}\widehat{\bm C}^{\bm\top}\bm w_{i-1}\bm w_{i-1}^{\bm\top}\widehat{\bm C}\widehat{\bm C}^{\bm\top}}=\sum_{i=2}^{n}{\bm w_{i-1}\bm w_{i-1}^{\bm\top}}\\
&=\bm I_m-\sum_{i=n}^{m}{\bm w_i\bm w_i^{\bm\top}}.
\end{align*}
By Lemma \ref{lemma2}, $\bm w_i$ for $i=n,\cdots, m$ are the orthonormal eigenvectors corresponding 
to the zero eigenvalue such that $\bm w_i^{{\bm\top}}\widehat{\bm C}\widehat{\bm C}^{{\bm\top}}\bm w_i=0$
from which we obtain $\widehat{\bm C}^{{\bm\top}}\bm w_i=0$. Since $\widehat{\bm C}^{{\bm\top}}=\bm M^{-1/2}\widetilde{\bm C}\bm R^{1/2}$, $\widetilde{\bm C}\bm R^{1/2}\bm w_i=0$ which indicates that the vectors
$\bm w_i$ for $i=n,\cdots,m$ form an orthonormal basis of the kernel of $\widetilde{\bm C}\bm R^{1/2}$.  Define
$\bm X_i=\bm w_{i+n-1}$ for $i=1,\cdots, m-n+1$ to complete the proof. \hfill $\square$
\par 
\begin{corollary}\label{corollary4}
If Assumption \ref{assumption} holds and $l_{c_{ij}}=\gamma$ for all the lines, the variance matrix $\bm Q_\delta$ becomes
\begin{align}\label{variancecorollary}
\bm Q_\delta=\frac{\eta}{2\gamma}\big(\bm I_m-\sum_{i=1}^{m-n+1}\bm X_i\bm X_i^{\bm\top}\big)
\end{align}
where $\bm X_i$ becomes the orthonormal basis of the kernel of the incidence matrix $\widetilde{\bm C}$. 
Furthermore, the trace of $\bm Q_\delta$ is 
$\ds\frac{\eta}{2\gamma}(n-1)$.
\end{corollary}
The proof follows directly from Theorem \ref{theoremmain} with $\bm R=\gamma \bm I_m$ and 
\begin{align*}
 \text{trace}\big(\ds\sum_{i=1}^{m-n+1}\bm X_i\bm X_i^{{\bm\top}}\big)=\sum_{i=1}^{m-n+1}\bm X_i^{\bm\top}\bm X_i=m-n+1.
\end{align*}
The trace of $\bm Q_\delta$ has been obtained by the $\mathcal{H}^2$ norm of input-output linear systems as in \cite{optimal_inertia_placement,H2norm}, which
is consistent with the result in the above corollary. 
\par 
Following the procedure described in Appendix \ref{kernelspace}, the vector $\bm X_i$ can be calculated from the basis vectors of the kernel of $\widetilde{\bm C}\bm R^{1/2}$.  
Due to the non-uniqueness of the basis vectors $\bm{\xi}_c$ for $c=1,\cdots,m-n+1$ of the kernel of $\bm C$, the set of the orthonormal 
basis vectors of the kernel of $\bm C\bm R^{1/2}$ is also non-unique. However, for the kernel, a set of orthonormal basis vectors can be obtained
from any set of basis vectors by a linear transformation consisting of an orthogonal matrix. Such a
transformation does not influence the calculation of the multiplication $\bm X_i\bm X_i^{\bm\top}$. 
The explicit formula (\ref{theoremmain-1}) of $\bm Q_\delta$ describes the dependence of the variances of the phase angle differences on the system parameters.  It is shown that the variances of the phase angle differences increase linearly as the disturbance-damping ratio $\eta$ increases.  
Because the variance of the phase-angle differences does not depend on the inertia, the control objective
of rotor angle stability hardly be improved by changing the virtual inertia. Here, the rotor angle 
stability is the ability of the phase angles to maintain their coherence. 
\par
In particular, formula (\ref{theoremmain-1}) reveals the role of the network topology with weight $l_{c_{ij}}$ for line $e_k$. In the complex network theory, the kernel of $\widetilde{\bm C}$ is the cycle space of the graph $\mathcal{G}$. Hence, \emph{it follows from formula (\ref{theoremmain-1}) that the stability of the power system is related to the cycle space of the graph. The way that changes in the topology of the power network affect the variances of the phase angle differences and hence stability can be investigated by a study of the cycle space of the graph}. 
 In Section \ref{Section:network}, we make a further study on the impact of the network topology by studying the cycle space of graphs. 

\section{Bounds of the variance matrices for networks with non-uniform disturbance-damping ratio}\label{Section:bound}
In the previous sections, we discussed the roles of the parameters 
in systems with a uniform disturbance-damping ratio at the nodes.
In this section, we present the findings for a system with non-uniform ratios. 
We define $\overline{\eta}=\max\{\eta_i,i=1,\cdots,n\}$ and $\underline{\eta}=\min\{\eta_i,i=1,\cdots,n\}$ 
with $\eta_i=b_i^2/d_i$. For $\bm A,\bm B\in\mathbb{R}^{n\times n}$, we say that $\bm A\leq \bm B$ 
if the matrix $\bm A-\bm B$ is semi-negative-definite.

\begin{lemma}\label{sec6-lemma1}
Define $\overline{\eta}=\max\{\eta_i,i=1,\cdots,n\}$ and $\underline{\eta}=\min\{\eta_i,i=1,\cdots,n\}$ 
with $\eta_i=b_i^2/d_i$,  and define $\overline{\bm \beta}=(\overline{\eta}\bm D)^{1/2}$ 
and $\underline{\bm\beta}=(\underline{\eta}\bm D)^{1/2}$. 
The solution $\bm Q_x$ of the Lyapunov equation (\ref{barQ}) satisfies 
the following inequalities where the various matrices are also defined
\begin{align}\label{sec6-2}
\bm Q_{\underline{\beta}}\leq \bm Q_x\leq \bm Q_{\overline{\beta}},
\end{align}
where 
\begin{align*}
\bm Q_{\overline{\beta}}=\int_0^{\infty}e^{\bm A_2 t}\overline{\bm B}_2\overline{\bm B}_2^{{\bm\top}}e^{\bm A_2^{{\bm\top}}t}\emph{d}t,~
\bm Q_{\underline{\beta}}=\int_0^{\infty}e^{\bm A_2 t}\underline{\bm B}_2\underline{\bm B}_2^{{\bm\top}}e^{\bm A_2^{{\bm\top}}t}\emph{d}t
\end{align*}
with $\overline{\bm B}_2,\underline{\bm B}_2\in\mathbb{R}^{(2n-1)\times n}$ such that
\begin{align*}
\overline{\bm B}_2=
\begin{bmatrix}
\bm 0\\
\bm U^{{\bm\top}}\bm M^{-1/2}\overline{\bm \beta}
\end{bmatrix},~~
\underline{\bm B}_2=
\begin{bmatrix}
\bm 0\\
\bm U^{{\bm\top}}\bm M^{-1/2}\underline{\bm \beta}
\end{bmatrix}.
\end{align*}
\end{lemma}
\emph{Proof:} By the definition of $\overline{\bm \beta}$ and 
$\underline{\bm \beta}$ and $\underline{\eta}d_i\leq b_i^2=\eta_i d_i\leq \overline{\eta}d_i)$ for all the nodes,
we obtain 
\begin{align*}
\underline{\eta}\text{diag}(d_i)=\underline{\bm \beta}\underline{\bm \beta}^{\bm\top}\leq \widetilde{\bm B}\widetilde{\bm B}^{\bm\top}=\text{diag}(b_i^2) \leq \overline{\bm \beta}\overline{\bm \beta}^{\bm\top}=\overline{\eta}\text{diag}(d_i).
\end{align*}
Hence, with the definition of $\bm B_2$ in (\ref{matrix-A2})
\begin{align*}
\underline{\bm B}_2\underline{\bm B}_2^{\bm\top}\leq \bm B_2\bm B_2^{{\bm\top}}\leq \overline{\bm B}_2\overline{\bm B}_2^{\bm\top}
\end{align*}
which leads to (\ref{sec6-2}).  \hfill $\square$
\par 
Based on Lemma \ref{sec6-lemma1}, we deduce bounds for $\bm Q_\omega$ and $\bm Q_\delta$. 
\begin{theorem}\label{theorem4}
Consider the system (\ref{stochasticsystem}). 
The variance matrix $\bm Q_\omega$ of the frequencies at the nodes 
satisfies 
\begin{align}\label{theorem4-1}
\frac{1}{2}\underline{\eta}\bm M^{-1}\leq\bm Q_\omega\leq \frac{1}{2}\overline{\eta}\bm M^{-1},
\end{align}
the variance matrix $\bm Q_\delta$ 
of the phase angle differences in the lines satisfies 
\begin{align}\label{theorem4-2}
\frac{1}{2}\underline{\eta}
\widehat{\bm Q}\leq \bm Q_\delta
\leq \frac{1}{2}\overline{\eta}\widehat{\bm Q}, ~~\widehat{\bm Q}=
\bm R^{-1/2}\big(\bm I_m-\sum_{i=1}^{m-n+1}\bm X_i\bm X_i^{{\bm\top}}\big)\bm R^{-1/2},
\end{align}
where $\underline{\eta}$ and $\overline{\eta}$
are defined in Lemma \ref{sec6-lemma1} and $\bm X_i$ is as defined 
in Theorem \ref{theoremmain}. 
\end{theorem}
\emph{Proof:} In Lemma \ref{sec6-lemma1}, the matrices $\underline{\bm B}_2$ and 
$\overline{\bm B}_2$ are defined such that the disturbance-damping ratio $b_i^2/d_i=\underline{\eta}$ and $b_i^2/d_i=\overline{\eta}$  for all the nodes respectively. 
Hence, using Lemma \ref{lemma1}, $\bm Q_{\underline{\beta}}$ and $\bm Q_{\overline{\beta}}$ are solved 
explicitly  as 
\[
\bm Q_{\underline{\beta}}=
\begin{bmatrix}
\frac{1}{2}\underline{\eta}\bm \Lambda_{n-1}^{-1}&\bm 0\\
\bm 0&\frac{1}{2}\underline{\eta}\bm I
\end{bmatrix}
, ~~
\bm Q_{\overline{\beta}}=
\begin{bmatrix}
\frac{1}{2}\overline{\eta}\bm \Lambda_{n-1}^{-1}&\bm 0\\
\bm 0&\frac{1}{2}\overline{\eta}\bm I
\end{bmatrix}.
\]

From (\ref{sec6-2}), we obtain
\begin{align}\label{theorem4-proof-2}
\bm C_2 \bm Q_{\underline{\beta}}\bm C_2^{\bm\top}\leq \bm Q=\bm C_2 \bm Q_x\bm C_2^{\bm\top}\leq \bm C_2 \bm Q_{\overline{\beta}}\bm C_2^{\bm\top},
\end{align}
where $\bm C_2$ is the one in (\ref{C2-delta-omega}) or in (\ref{C2-delta}) or in (\ref{C2-omega}).  
\par 
To prove (\ref{theorem4-1}), we consider the
output as the frequency and take $\bm C_2$ in (\ref{C2-omega}).
Following 
the procedure to calculate the variances
of the frequencies in Theorem \ref{theorem-omega} with $b_i^2/d_i=\overline{\eta}$ and $b_i^2/d_i=\underline{\eta}$ for all the nodes, we 
get 
\begin{align*}
\bm C_2 \bm Q_{\underline{\beta}}\bm C_2^{\bm\top}= \frac{1}{2}\underline{\eta}\bm M^{-1}~~\text{and}~~\bm C_2 \bm Q_{\overline{\beta}}\bm C_2^{\bm\top}= \frac{1}{2}\overline{\eta}\bm M^{-1},
\end{align*}
which lead to (\ref{theorem4-1}) with (\ref{theorem4-proof-2}). 
\par 
To prove (\ref{theorem4-2}), we consider the 
output as the phase angle differences and insert $\bm C_2$ of (\ref{C2-delta})
into (\ref{theorem4-proof-2}), then obtain the upper bound of $\bm Q_\delta$ from (\ref{proof-theoremmain-0}) such that 
\begin{align*}
\hspace{-15pt}
\bm C_2\bm Q_{\overline{\beta}}\bm C_2^{\bm\top}
=\frac{1}{2}\overline{\eta}\widetilde{\bm C}^{\bm\top}\bm M^{-1/2}(\bm M^{-1/2}\bm L_c \bm M^{-1/2})^{\dag}\bm M^{-1/2}\widetilde{\bm C}.
\end{align*}
Following the procedure to deduce the explicit formula in (\ref{proof-theoremmain-1}), we obtain 
\begin{align*}
\widetilde{\bm C}^{\bm\top}\bm M^{-1/2}(\bm M^{-1/2}\bm L_c \bm M^{-1/2})^{\dag}\bm M^{-1/2}\widetilde{\bm C}=\widehat{\bm Q}.
\end{align*}
Hence, the upper bound of $\bm Q_\delta$ satisfies
$\bm C_2\bm Q_{\overline{\beta}}\bm C_2^{\bm\top}=\frac{1}{2}\overline{\eta}\widehat{\bm Q}$.
Similarly,  the lower bound satisfies $\bm C_2\bm Q_{\underline{\beta}}\bm C_2^{\bm\top}= \frac{1}{2}\underline{\eta}\widehat{\bm Q}$.
With these two bounds and (\ref{theorem4-proof-2}), we obtain (\ref{theorem4-2}). \hfill $\square$
\par 
It is well known that the diagonal elements of a semi-positive definite symmetric 
matrix are all non-negative. Hence, the bounds of the variances 
of the frequencies at the nodes and the phase angle differences in the lines 
are derived directly from (\ref{theorem4-1}) and (\ref{theorem4-2}).  
\par 
Formula (\ref{theorem4-1}) reveals the factors that impact the variances of the
frequencies at nodes in networks with a non-uniform disturbance-damping ratio. First, as in networks with a uniform disturbance-damping ratio, the inertias of the synchronous machines locally impact the variances
of the frequencies at the nodes, and the network topology and the parameter $l_{c_{ij}}$
have little impact because they are absent in the formula. Second, in networks with a non-uniform disturbance-damping ratio, the variances of the frequencies will increase
as the minimum value $\underline{\eta}$ increases
and decrease as the maximum value $\overline{\eta}$ decreases. Hence,
by decreasing all the disturbance-damping ratios, the variances 
of the frequencies will be decreased, which is consistent with the findings in 
networks with a uniform disturbance-damping ratio. In addition, by decreasing 
the maximum value $\overline{\eta}$, there are nodes at which the variances of the frequencies
will be decreased. 
\par 
Formula (\ref{theorem4-2}) illustrates the roles played by the system parameters in determining 
the variances of the phase angle differences in networks with a non-uniform disturbance-damping ratio. 
First, the roles of the values $\underline{\eta}$ and $\overline{\eta}$ in determining 
the variances of the phase angle differences is the same as that in determining the variances 
of the frequencies. Decreasing the largest disturbance-damping ratio can decrease the 
variances of the phase angle differences at some lines. 
For example, energy storage in combination with droop control, which affects
the parameter $d_i$ at the relevant nodes, will directly decrease the disturbance-damping ratios. Second, as in a network with a uniform disturbance-damping ratio, 
the inertia is absent from the formula, and the role of the network topology is also reflected by the basis of the cycle space. Hence, the inertia has little impact on the variances of the phase angle differences, and by forming small cycles, the variances of the phase angle differences can also be effectively decreased 
in the network. Third, the impact 
of constructing new lines to form cycles and increasing the
capacities of the lines on the upper and lower bounds are the 
same as in the networks with a uniform disturbance-damping ratio.

In regard to the impact of the scales of the power systems on the stability, we have the following conclusion. 
From formulas (\ref{theorem0-1},\ref{theoremmain-1},~\ref{theorem4-1},\ref{theorem4-2}), we see that, if the scale of the network is increased by constructing nodes that have small effects on the power flows and possess disturbance-damping ratios close to $\eta$, the fluctuations in the frequency or in the phase angle differences in the network will not be dramatically increased or decreased. Hence, the stability will be changed little by increasing the scale of the network. 
This follows formula (\ref{theorem0-1}) for networks with a uniform disturbance-damping ratio, which states that the newly connected nodes with disturbance-damping ratios equal to $\eta$ will not bring fluctuations to the frequency at the other nodes. Since $\delta_i^*\approx \delta_j^*$ for all the nodes, the newly connected nodes have little influence on the phase angle difference in the synchronous state, and it is indicated by formula (\ref{theoremmain-1}) that the fluctuation of the phase angle difference will not change greatly. Similarly,
for networks with a non-uniform disturbance-damping ratio, the newly connected nodes with disturbance-damping ratios in the set $[\underline{\eta},\overline{\eta}]$ will not change the bounds of the variance, 
as follows from the formulas (\ref{theorem4-1}) and (\ref{theorem4-2}).
This conclusion is different from that obtained by a study of linear stability \citep{Xi2016}, where the
linear stability decreases if the scale of the network increases. 
However, if nodes that consume a large amount of power and have large disturbance-damping ratios $b_i^2/d_i$ are added to the network, the variance of the frequency and the phase angle difference may increase because
the weights of lines may decrease and the disturbances 
may propagate from these nodes to the other nodes in the network.

\section{The role of the network topology}\label{Section:network}
To fully explore the role of the network topology from the formula (\ref{theoremmain-1}), we introduce three concepts for graphs,
\begin{definition}
Consider a connected and undirected graph $\mathcal{G}$. (i) A single line is defined as a line that does not belong to any cycle; (ii) Line $e_1$ is called  a cycle-shared line of line $e_2$ if there exists at least one cycle containing both $e_1$ and $e_2$; (iii) A cycle-cluster is a subgraph of $\mathcal{G}$ obtained in the following way. One starts from
a subgraph of one cycle and extends it by adding the lines in all the cycles with which the subgraph 
has at least one line in common, then one obtains a cycle-cluster.  
\end{definition}
It is deduced that a graph is composed of cycle-clusters and single lines, a line either belongs to a cycle-cluster or is a single line and in a cycle-cluster each pair of lines are cycle-shared lines.
In the following example, we explain the definitions and the formulation of the basis vectors of the cycle space. 
\begin{figure}[h]
\centering
\includegraphics[scale=.06]{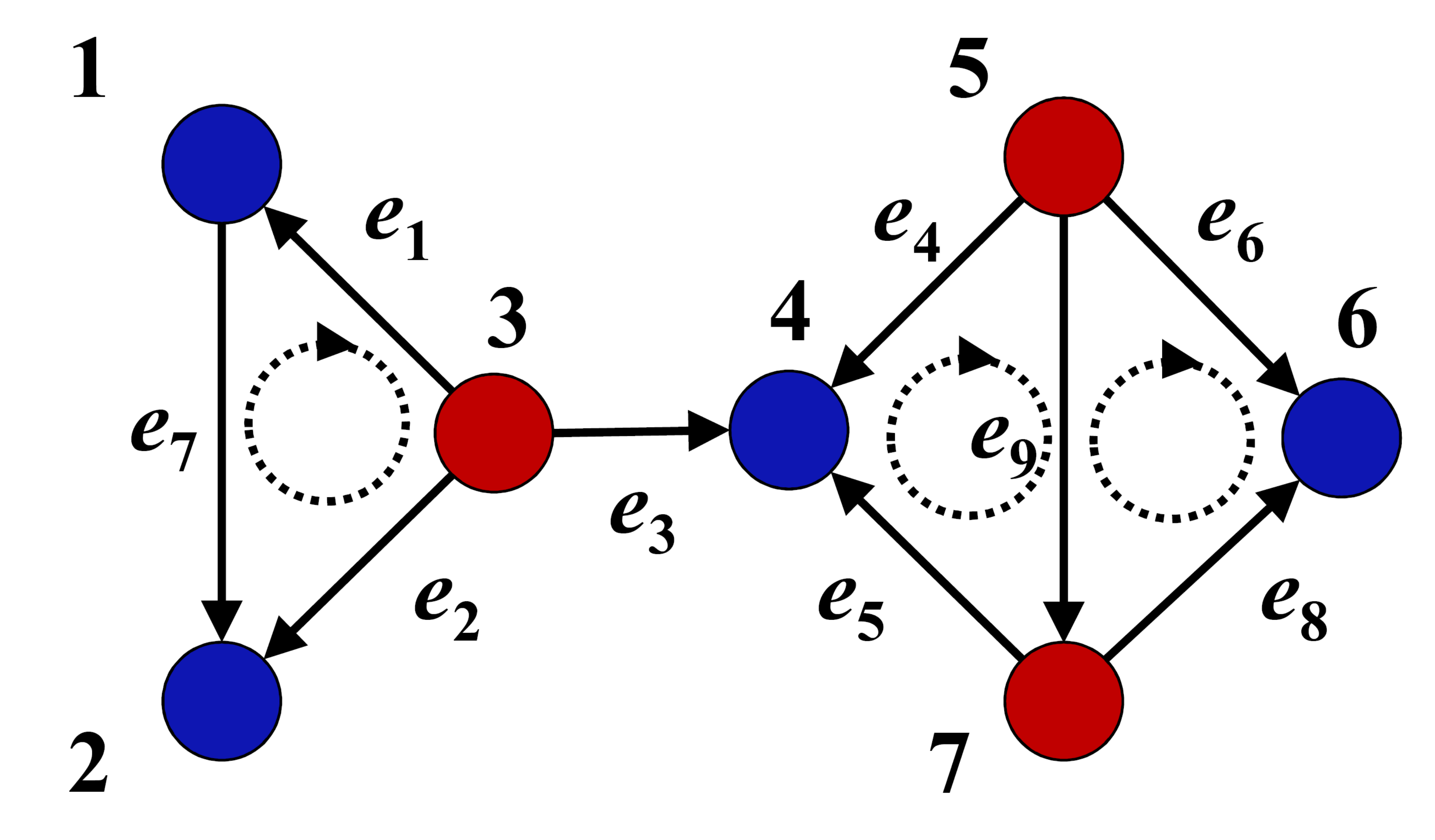}
\caption{A network with two cycle-clusters and a single line.\label{fig0}}
\end{figure}
\begin{example}
Consider the network show in Fig. \ref{fig0}.  There are two cycle-clusters, i.e.,  $\{e_1,e_2,e_7\}$ and $\{e_4,e_5,e_6,e_8,e_9\}$, and a single line $e_3$ that does not belong to any cycle.  Each pairs of lines 
in the cycle-cluster $\{e_1,e_2,e_7\}$ are cycle-shared respectively, similarly for the lines in the cycle-cluster 
$\{e_4,e_5,e_6,e_8,e_9\}$. However, two lines belonging to two different cycle-clusters are not cycle-shared, because a cycle containing both of these two lines cannot be found, for example $e_1$ and $e_4$. 
The directions of lines are specified for the formulation of the incidence matrix $\widetilde{\bm C}$ and
the calculation of the basis vectors of the cycle space. The directions of all
the cycles are clock-wise. 
Following the procedure to calculate 
the basis vectors of the cycle space in Appendix \ref{kernelspace}, we get the basis vectors of the cycle space of this network,
$\bm\xi_1=\begin{bmatrix}
-1&1&0&0&0&0&-1&0&0
\end{bmatrix}^{\bm \top}$,
$\bm \xi_2=\begin{bmatrix}
0&0&0&-1&1&0&0&0&1
\end{bmatrix}^{\bm \top}$,
and 
$\bm \xi_3=\begin{bmatrix}
0&0&0&0&0&1&0&-1&-1
\end{bmatrix}^{\bm \top}$
which are corresponding to the fundamental cycles $\{e_1,e_2,e_7\}$, $\{e_4,e_5,e_9\}$ and $\{e_6,e_8,e_9\}$ respectively. Obviously, $\bm \xi_1$ is 
orthogonal to $\bm \xi_2$ and $\bm \xi_3$. This indicates that the basis vectors corresponding 
the cycles in different cycle-clusters are orthogonal. 
Due to the non-uniqueness of the spanning tree selected to form the fundamental cycles,
the basis vectors are also non-unique.  Thus
the basis vectors of the cycle space of the network in Fig. \ref{fig0} can also be 
$\bm\xi_1=\begin{bmatrix}
-1&1&0&0&0&0&-1&0&0
\end{bmatrix}^{\bm \top}$,
$\bm \xi_2=\begin{bmatrix}
0&0&0&-1&1&0&0&0&1
\end{bmatrix}^{\bm \top}$,
and 
$\bm \xi_3=\begin{bmatrix}
0&0&0&-1&1&1&0&-1&0
\end{bmatrix}^{\bm \top}$
which are corresponding to the fundamental cycles $\{e_1,e_2,e_7\}$, $\{e_4,e_5,e_9\}$ and $\{e_4,e_6,e_8,e_5\}$ respectively. 
\end{example}
\par
The network topology has two effects on the stability of the power system: the power flows at the synchronous state $(\bm \delta^*,\bm 0)$ and the variance of the phase angle differences. Formula (\ref{theoremmain-1}) indicates that the variance also depends on  the power flows because $R_k=l_{c_{ij}}$ and $l_{c_{ij}}=l_{ij}\cos{\delta_{ij}^*}$. This demonstrates the nonlinear character of the impacts of the network topology on stability. A network can be constructed mathematically in two steps, i.e., first connecting all the nodes to form a tree network and then constructing new lines or replacing the existing lines by ones with larger capacities. By following these steps, in addition to investigating the tree network, we reveal the role of the network topology by studying the impact of constructing new lines and increasing the capacity of the lines. 
\par 
For the power flows, we have the 
following proposition.
\begin{proposition}\label{proposition-powerflow}
Consider the power system (\ref{Nonlinear}) with a synchronous state that satisfies 
the security condition (\ref{condition}). (i) If the capacity of a single line is increased, then the power flows in all the other lines remain unchanged. (ii) If in  a cycle-cluster a new line is constructed 
or the capacity of a line is increased, the power flows in the lines
that are not in this cycle-cluster, remain unchanged. 
\end{proposition}
\emph{Proof:} Without loss of generality, we assume there three sub-graphs in graph $\mathcal{G}$, i.e., 
$\mathcal{G}_1(\mathcal{V}_1,\mathcal{E}_1)$, $\mathcal{G}_2(\mathcal{V}_2,\mathcal{E}_2)$  and $\mathcal{G}_3(\mathcal{V}_3,\mathcal{E}_3)$ where $\mathcal{G}_1(\mathcal{V}_1,\mathcal{E}_2)$ is either a cycle-cluster or single-line, 
$\mathcal{V}_1\cup\mathcal{V}_2\cup\mathcal{V}_3=\mathcal{V}$, $\mathcal{E}_1\cup\mathcal{E}_2\cup\mathcal{E}_3=\mathcal{E}$, $\mathcal{E}_i\cap\mathcal{E}_j=\emptyset$ for $i\neq j$, $\mathcal{V}_1\cap\mathcal{V}_2=\{k\}$,
$\mathcal{V}_1\cap\mathcal{V}_3=\{q\}$ and $\mathcal{V}_2\cap\mathcal{V}_3=\emptyset$.
We prove that the power flows in the lines in $\mathcal{G}_2$ remain unchanged 
when the capacity of a line is increased or a new line is constructed in $\mathcal{G}_1$.
In the power flow calculation, we choose node $k$ as the reference node with $\delta_k=0$. Thus, 
the power flow in the $\mathcal{G}_1$ and $\mathcal{G}_2$ are decoupled, where the power flows in cycle-cluster $\mathcal{G}_2$ satisfy
\begin{align*}
P_i-\sum_{j\in\mathcal{V}_2}l_{ij}\sin{(\delta_i-\delta_j)}&=0, ~i\in\mathcal{V}_2~~\text{and}~~i \neq k, \\
\delta_k&=0.
\end{align*}
that are not changed by adding new lines or increasing the capacities of lines in cycle-cluster $\mathcal{G}_1$. Similarly, it is proven that the power flows in $\mathcal{G}_3$ remain unchanged by 
constructing new lines or increasing the capacity of the lines in $\mathcal{G}_1$.  
 \hfill $\square$


 \par 
From Proposition \ref{proposition-powerflow}, we obtain that the phase angle 
differences $\delta_{ij}^*$ in lines at the synchronous state in a cycle-cluster are independent of the power flows
in the other cycle-clusters. Hence, the weights $l_{c_{ij}}=l_{ij}\cos{\delta_{ij}^*}$ of lines 
in the cycle-cluster will not be changed by adding lines or increasing line capacity 
in the other cycle-clusters. 
\par 
Based on the theory of the cycle space,  we obtain the following Corollary of Theorem \ref{theoremmain}.
\begin{corollary}\label{main-corollary1}
Consider the system (\ref{stochasticsystem}) with Assumption \ref{assumption}.
\begin{enumerate}[(i)]
\item The invariant probability distribution of the phase angle difference in a single line connecting nodes $i$ and $j$ is independent of
those of the phase angle differences in all the other lines in the network,
and the variance of the phase angle differences in this line is  $\ds \frac{1}{2}\eta l_{c_{ij}}^{-1}$.
\item According to the invariant probability distribution, the phase-angle differences of all lines in a particular
cycle cluster are independent of the phase-angle differences of all lines which are not in this cycle
cluster.
\item Increasing the weight of a line or constructing 
new lines in a cycle-cluster without changing the weights of all the 
other lines decreases the variances of the phase angle differences in the lines of this cycle-cluster.
\item For a cycle-cluster with only one cycle with lines in set $\mathcal{E}_c$ in the graph, the variance of the phase angle 
differences in the line connecting nodes $i$ and $j$ in this cycle-cluster is
\begin{align}\label{singlecycle}
\frac{\eta}{2}\Big(l_{c_{ij}}^{-1}-l_{c_{ij}}^{-2}\Big(\sum_{(r,q)\in\mathcal{E}_c}l_{c_{rq}}^{-1}\Big)^{-1}\Big).
\end{align} 
If $l_{c_{ij}}=\gamma$ for all the lines in this cycle, the variances of the phase angle differences in
these lines are $\ds\frac{\eta}{2\gamma}(1-\frac{1}{N})$ where $N$ is the
length of the cycle.
\end{enumerate}
\end{corollary}
\emph{Proof:} (i) For an acyclic network, it follows from (\ref{theoremmain-1})
that the variance matrix of the phase angle difference is $\ds\frac{\eta}{2}\bm R^{-1}$
because the cycle space of the acyclic network is empty. Thus, the variance in line $e_k$ is 
$\ds\frac{1}{2}\eta l_{c_{ij}}^{-1}$. For a network with cycles and single lines, without loss of generality, assume line $e_1$ is a single-line. Following 
the method to formulate the basis of the cycle space in Appendix \ref{kernelspace}, the base vector 
has the form 
$\bm \xi_i=\begin{bmatrix}
0&\xi_{i,2}&\xi_{i,3}&\cdots&\xi_{i,m}
\end{bmatrix}^{\bm\top}$
where $\xi_{i,j}$ is either $-1$, $1$ or $0$, and $\bm X_i$ 
has the form 
$\bm X_i=
\begin{bmatrix}
0&x_{i,2}&x_{i,3}&\cdots&x_{i,m}
\end{bmatrix}
^{\bm\top}$ obtained by Gram-Schmidt 
orthogonalization of $\bm R^{-1/2}\bm \xi_i$. 
Because the elements in the first column and the first row of $\bm X_i\bm X_i^{\bm\top}$ 
are all zero,  we derive the independence 
of the invariant probability distribution of the phase angle difference in this line to
those of the phase angle difference in all the other lines. By (\ref{theoremmain-1}), 
we obtain that the variance in this line is $\ds \frac{\eta}{2l_{c_{ij}}}$. 
\par 
(ii) We partition the graph $\mathcal{G}$ into two sub-graphs, $\mathcal{G}_1$ and $\mathcal{G}_2$,
where $\mathcal{G}_1$ is either a cycle-cluster or a single line. If $\mathcal{G}_1$ is 
a single line, we obtain this conclusion directly from Corollary \ref{main-corollary1}(i) directly. 
We now consider the case where $\mathcal{G}_1$ is a cycle-cluster. 
Denote the number
of lines in these two sub-graphs by $N$ and $m-N$, the number of fundamental cycles by $m_1$ and $m_2$, the lines in $\mathcal{G}_1$ by
$e_1,\cdots,e_N$ and those in $\mathcal{G}_2$ by $e_{N+1},\cdots,e_m$ respectively. 
Here, $m_1+m_2=m-n+1$.
The basis vectors of the cycles in $\mathcal{G}_1$ have the form 
$\bm \xi_i=
\begin{bmatrix}
\xi_{i,1}&\xi_{i,2}&\cdots&\xi_{i,N}&0&\cdots&0
\end{bmatrix}^{\bm\top}$ for $i=1,\cdots,m_1$ and 
those of the cycles in  $\mathcal{G}_2$ have the form $\bm \xi_i=
\begin{bmatrix}
0&0&\cdots&0&\xi_{i,{N+1}}&\cdots&\xi_{i,m}
\end{bmatrix}^{\bm\top}$ for $i=m_1+1,\cdots,m-n+1$.
In these vectors, $\xi_{i,j}$ are either $1$,$-1$, or $0$. 
By Gram-Schmidt orthogonalization of $\bm R^{-1/2}\bm \xi_i$, we get the orthonormal
vectors $\bm X_i=
\begin{bmatrix}
x_{i,1}&\cdots&x_{i,N}&0&\cdots&0
\end{bmatrix}^{\bm\top}$ for 
$i=1,\cdots,m_1$ and $\bm X_i=
\begin{bmatrix}
0&\cdots&0&x_{i,{N+1}}&\cdots&x_{i,m}
\end{bmatrix}^{\bm\top}$ for 
$i=m_1+1,\cdots,m-n+1$.  
It is obvious that the entries in the first $N$ columns and the first $N$ rows of the matrix $\ds\sum_{i=m_1+1}^{m-n+1}\bm X_i\bm X_i^{\bm\top}$ 
are all $0$.  This indicates that the lines in $\mathcal{G}_2$ have no contributions to 
 the first $N$ columns and the first $N$ rows of $\bm Q_\delta$.  Similarly, the lines in
 $\mathcal{G}_1$ have no contributions to the last $m-N$ columns and the last $m-N$ rows of $\bm Q_\delta$. Hence, 
 the invariant probability distribution of the phase angle differences in the lines of $\mathcal{G}_1$ are 
 independent of those in the lines of $\mathcal{G}_2$. 
\par 
(iii) The case in which the weight 
of a line in a cycle-cluster increases is considered first. 
Assume the graph is a cycle-cluster, where the weight of line $e_1$ increases. 
Denote the dimension of the kernel of 
$\widetilde{\bm C}\bm R^{1/2}$ by $N$, which equals to $m-n+1$. Thus, there are $N$ fundamental cycles
in the cycle-cluster. The basis vectors are chosen below. 
The basis vectors corresponding to the $N-1$ fundamental cycles which 
do not include line $e_1$ 
have the form 
$\bm\xi_i=
\begin{bmatrix}
0&\xi_{i,2}&\xi_{i,3}&\cdots&\xi_{i,m}
\end{bmatrix}^{\bm\top}
$ for $i=1,\cdots,N-1$, where $\xi_{i,q}=1,-1~\text{or}~0$ for $q=2,\cdots, m$ and that corresponding to the fundamental cycle which includes line $e_1$ has the form 
$\bm \xi_N=
\begin{bmatrix}
\xi_{N,1}&\xi_{N,2}&\cdots&\xi_{N,m}
\end{bmatrix}^{\bm\top}
$ where $\xi_{N,1}=1~\text{or}~-1$ and 
$\xi_{N,q}=1,-1~\text{or}~0$ for $q=2,\cdots,m$. 
This can be done by changing the basis vectors of the cycle space properly. 
By the Gram-Schmidt orthogonalization of $\bm R^{-1/2}\bm \xi_i$,
we obtain $\bm X_i=
\begin{bmatrix}
0&x_{i,2}&x_{i,3}&\cdots&x_{i,m}
\end{bmatrix}^{\bm\top}
$ for $i=1,\cdots,N-1$ which is independent of the weight $l_1$ of line $e_1$.
The last unit vector 
$\bm X_N$ can be obtained by the normalization of the vector $\bm X_N'=\bm R^{-1/2}\bm\xi_N-\alpha_1\bm X_1-\cdots-\alpha_{N-1}\bm X_{N-1}$ where 
$\ds\alpha_i=\frac{(\bm R^{-1/2}\bm\xi_N)^{\bm \top}\bm X_i}{\bm X_i^{\bm \top}\bm X_i}$.
Because the first element of $\bm X_i$ is zero for $i=1,\cdots,N-1$, $\alpha_i$ is independent of $l_1$.
Hence $\bm X_N'$
has the form $\bm X_N'=
\begin{bmatrix}
l_1^{-1/2}\xi_{N,1}&x'_{N,2}&x'_{N,3}&\cdots&x'_{N,m}
\end{bmatrix}^{\bm\top}
$ where $x'_{N,q}$ is independent of $l_1$ for $q=2,\cdots,m$. By the normalization of $\bm X_N'$,
we obtain $\bm X_N=a\bm X_N'$ where 
$a=\big(l_1^{-1}+\ds\sum_{i=2}^{m}{x'_{N,i}}^2\big)^{-1/2}$. Hence, 
the diagonal element of $\bm X_N\bm X_N^{\top}$ equals to   
$a^2l_1^{-1}$ for $i=1$ and equals to $a^2{x'_{N,i}}^2$ for $i=2,\cdots,m$. 
Inserting $\bm X_N\bm X_N^{\bm \top}$ into 
(\ref{variancecorollary}), we obtain
the variance of the phase angle difference in line $e_1$ which equals to $\ds\frac{1}{2}\eta(l_1^{-1}-a^2l_1^{-2})$ and that in line $e_q$ which equals to $\ds\frac{1}{2}\eta l_q^{-1}(1-a^2{x'_{N,q}}^2-\sum_{i=1}^{N-1}{x_{i,q}^2})$ for $q=2,\cdots,m$. It is 
obvious that if $l_1$ increases, these variances decrease.
\par 
 We next consider the case when a new line 
is constructed in a cycle-cluster without changing the weight of all the other lines. Assume line $e_1$ is the new line. 
Following the above calculation, we obtain that the 
variance in the line with weight $l_q$ equals to 
$\ds\frac{1}{2}\eta l_q^{-1}(1-a^2{x'_{N,q}}^2-\sum_{i=1}^{N-1}{x_{i,q}^2})$. 
For the variances in lines before constructing line $e_1$, by choosing the basis vector corresponding to the $N-1$ fundamental cycles which do not include line $e_1$ and 
the Gram-Schmidt orthogonalization of these vectors, we obtain the variance
in line $e_q$ with weight $l_q$  is 
$\ds\frac{1}{2}\eta l_q^{-1}(1-\sum_{i=1}^{N-1}{x_{i,q}^2})$ for $q=2,\cdots,m$.
 Clearly, the variance decreases after adding line $e_1$. 
\par 
 (iv) The lines in the cycle are denoted by $e_1,e_2,\cdots,e_N$ with weights $l_1,l_2,\cdots,l_N$. Assume the direction
of these lines are consistent with the direction of the cycle. The vectors corresponding 
to this cycle and the other cycles are denoted by $\bm \xi_1$ and $\bm \xi_i$ with $i=2,\cdots,m-n+1$ respectively.
Following Appendix \ref{kernelspace}, we obtain $\bm \xi_1=
\begin{bmatrix}
1&1&\cdots& 1&0&\cdots&0
\end{bmatrix}
^{\bm\top}$ where 
the first $N$ elements equal to $1$ and the last $m-N$ elements 
equal to $0$, and 
$\bm \xi _i=
\begin{bmatrix}
0&0&\cdots&0&\xi_{i,{N+1}}&\cdots&\xi_{i,m}
\end{bmatrix}^{\bm\top}
$ where the first $N$ elements
are all $0$ and the last $m-N$ elements equal to either $0$, $1$ or $-1$. Obviously, the vector $\bm R^{-1/2}\bm \xi_1$ is orthogonal 
to the vector $\bm R^{-1/2}\bm \xi_i$ for $i=2,\cdots,m-n+1$. By Gram-Schmidt 
orthogonalization, we derive $\bm X_1=
\Big(\sum_{k=1}^N l_{k}^{-1}\Big)^{-1/2}
\begin{bmatrix}
l_1^{-1/2}&l_2^{-1/2}&\cdots&l_N^{-1/2}&0&\cdots&0
\end{bmatrix}^{\bm\top}$ from $\bm R^{-1/2}\bm \xi_1$ and
$\bm X_i=
\begin{bmatrix}
0&0&\cdots&0&x_{i,{N+1}}&\cdots&x_{i,m}
\end{bmatrix}^{\bm\top}
$ for the linear subspace composed of the vectors $\bm R^{-1/2}\bm \xi_i$ with
$i=2,\cdots,m-n+1$. Because the first $N$ elements of $\bm X_i$ for $i=2,\cdots,m-n+1$ are all $0$, 
the matrix $\bm X_i\bm X_i^{\bm\top}$ has no contributions to the first $N$ columns 
and the first $N$ rows of $\bm Q_\delta$. Hence, the invariant probability distribution
of the phase angle differences in the lines of the cycle are independent from those in the other lines. Further more, by (\ref{theoremmain-1}), we obtain that the $k$th 
diagonal element of $\bm Q_\delta$ for $k=1,\cdots,N$ is
$$\frac{\eta}{2}\Big(l_{k}^{-1}-l_{k}^{-2}\big(\sum_{r=1}^N l_{r}^{-1}\big)^{-1}\Big)$$
from which we obtain (\ref{singlecycle}) by replacing $l_k$ by $l_{c_{ij}}$ for line $e_k$. If $l_k=\gamma$ for $k=1,\cdots,N$, we further get the first $N$ diagonal elements of $\bm Q_\delta$ equal to 
 $\ds\frac{\eta}{2\gamma}(1-\frac{1}{N})$. 
\hfill $\square$
\par

\begin{remark}\label{remarkfindings}
From Proposition \ref{proposition-powerflow} and Corollary \ref{main-corollary1}, we get the following findings. \\
\emph{(i) The variance of the phase angle difference in a single line connecting nodes $i$ and $j$ is  $\ds \frac{\eta}{2}l_{c_{ij}}^{-1}$, which is not influenced by either constructing a new line without forming a cycle-cluster that 
includes this line or increasing the capacities of the other lines. Thus, a single line is likely to
be a vulnerable line.} This is because neither the construction of new lines nor the increase
in the capacity of the other lines changes the power flow $l_{ij}\sin{\delta_{ij}^*}$ in this line, which 
is stated in  Proposition \ref{proposition-powerflow}, and the invariant probability distribution of the phase angle difference
in the single line is independent of those of the phase angle differences in all the other
lines, which is obtained from Corollary \ref{main-corollary1}-(i). \\ 
 \emph{(ii) Constructing new lines and increasing the capacities of lines in a cycle-cluster 
have no impact on the variances of the phase angle differences in the lines that are not in this cycle-cluster}.
This is because constructing new lines or increasing the capacities 
of lines in a cycle-cluster has no influence on the power flows in other cycle-clusters and single lines, which 
is indicated by Propostion \ref{proposition-powerflow}, and the invariant probability distribution of the phase angle differences in the lines of a cycle-cluster
is independent of those in the lines
that are not in this cycle-cluster, which is demonstrated by Corollary \ref{main-corollary1}-(ii). \\
 \emph{(iii) By either increasing the weights $l_{c_{ij}}$ of lines or constructing new lines without changing the weights of the other lines in a cycle-cluster,
the variances of the phase angle differences in this cycle-cluster will decrease.}
 \\
\emph{(iv) For a cycle-cluster with only one cycle with lines in set $\mathcal{E}_c$ in the graph, the variance of the phase angle 
difference in the line connecting nodes $i$ and $j$ can 
be calculated from (\ref{singlecycle}). In addition, based on (iii), we obtain that formula (\ref{singlecycle}) provides a conservative estimation of the variances in the lines in cycle-clusters, i.e., the variance in a line that is in multiple cycles can be approximated by formula (\ref{singlecycle})
by taking the smallest cycle that includes this line. }
\end{remark}
These findings provide guidelines on how to reduce the negative
effects of vulnerable lines and designing future power networks, which should have low variances in phase angle differences when subjected to stochastic disturbances from power sources
and power loads. The term \emph{remedy} will be used for the reduction of these negative
effects. 
Changing a power network
by adding lines to form small cycles or by increasing the capacity of particular lines
will suppress the fluctuations in the phase differences in the lines of the corresponding cycle-cluster.
The benefit of forming small cycles is that the fluctuations in the phase angle differences decrease by $O(1/N)$, where $N$ denotes the length of the cycle.
This is consistent with the findings obtained by studying 
the energy barrier of a nonlinear system with a cyclic network in Xi et al \citep{Xi2016}.
The fluctuations in the
phase angle differences can be decreased by replacing transmission lines with 
small line capacities by ones with large line capacities. This is the same rule as for the transient stability analysis of the \emph{Single Machine Infinite Bus} (SMIB) model \citep{Kundur} by the equal area criterion.
Because the variances of the phase angle differences decrease linearly with the parameter $l_{c_{ij}}=l_{ij}\cos{\delta_{ij}^*}$, the control of the power flows to increase 
the value $\cos{\delta_{ij}^*}$ can also decrease the fluctuations of the phase angle differences in 
the lines. These findings will be further explained in an example in Section \ref{casestudy}.
\par

\section{Case study}\label{casestudy}

\begin{figure}[h]
\includegraphics[scale=.115]{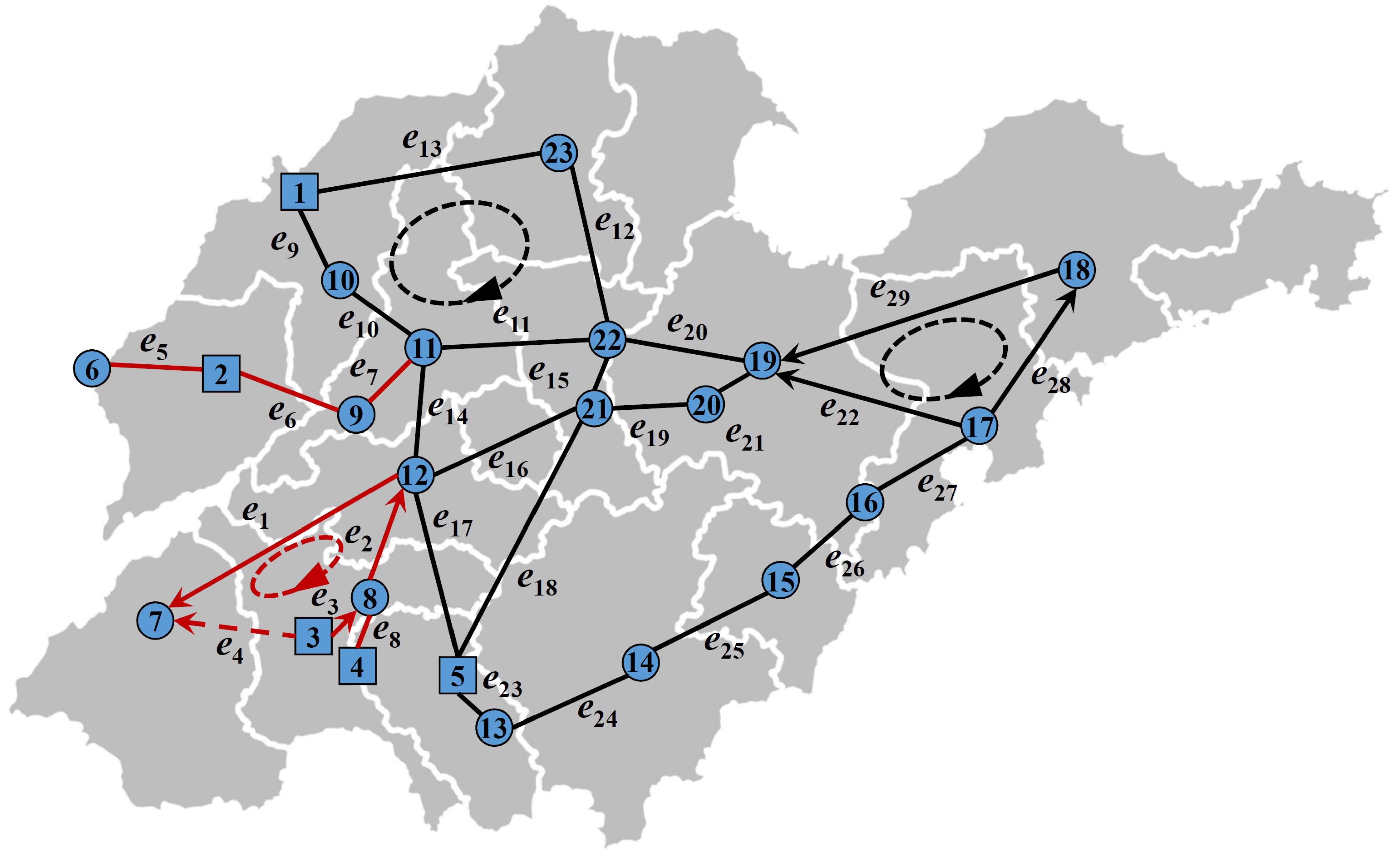}
\caption{500KV transmission network of Shandong province, China \label{fig1}}
\end{figure}

In this section, we verify the formulas (\ref{theorem0-1}, \ref{theoremmain-1}) for 
the networks with uniform disturbance-damping ratio, the bounds (\ref{theorem4-1}, \ref{theorem4-2}) for the 
variance matrices for the networks with non-uniform disturbance-damping ratio and 
the findings presented in Remark \ref{remarkfindings}. 
We take the 500KV transmission network of Shandong Province of China \citep{YEHUA} as an example. 
\begin{example}
Consider the 500 KV transmission network of Shandong Province as shown in Fig. \ref{fig1}.
There are 5 nodes with generators and 18 nodes with loads only. 
The nodes of squares denote power generators and the nodes of cycles denotes power loads.  Line $e_4$ does not exist in practice, which is constructed virtually in order to explain our findings. Before constructing line $e_4$, all the red lines are single lines and all 
the black lines are in a cycle-cluster which has 6 fundamental cycles. After constructing line $e_4$, there is one more cycle-cluster,
which is composed of $(e_1,e_2,e_3,e_4)$. We set $m_i=i$ for the generators and $m_i=1$ for all the 
loads. We study the 7 cases with different settings of $P_i$, $l_{ij}$ and $b_i^2/d_i$ as shown 
in Table \ref{table0}. 
\end{example}

\par 
In cases 1-3, it holds that the phase angle difference $\delta_{ij}^*=0$ because of $P_i=0$ for all the nodes. 
Thus, when disturbances occur, the frequencies at the nodes and the power flows in the lines fluctuate around zero. The weights of the lines satisfy $l_{c_{ij}}=l_{ij}$. 
The weights of the lines in Cases 4-6 are shown in Table \ref{table2}, which 
are calculated by solving the power flow equations. The variances of the frequencies at the nodes and the phase angle differences in the lines are presented in Tables \ref{table1} and \ref{table3}, respectively. The values in the tables 
are first calculated by formulas (\ref{theorem0-1}) and (\ref{theoremmain-1}) and 
then verified using Matlab following the procedure in Theorem \ref{lemmacovariance}. In Cases 1-3, because $\delta_{ij}^*=0$ for all the lines in the networks, the power flows are independent of the network topology.
In this case, the impact of the network topology alone on the variance of the phase angle difference 
can be observed. In Cases 4-6, because $P_i$ is nonzero, updating the network topology, such 
as constructing new lines and increasing the line capacities, may change the weight $l_{c_{ij}}$
or the cycle space. Hence, the overall impact of the network topology can be analysed. 
In case 3 and 6, the capacity of line $e_{23}$ is increased from $10$ to $20$ 
in order to observe the changes of the variances in the other lines. This line is selected 
because in Case 4 the variance in this line is the largest one 
in the cycle-cluster that includes this line, as shown in Table \ref{table3}. 
\par
First, let us focus on the variances of the phase angle differences in the single lines. 
Lines $e_1-e_8$ in cases 1, 4, and $e_5-e_8$ in cases 2-3,~5-6 are single lines. It is verified in Table \ref{table3} that the variances of the phase angle differences in these lines equal $\ds \frac{\eta}{2}l_{c_{ij}}^{-1}$ with the weights of the lines shown in Table \ref{table2}. In particular, the variances in lines $e_5-e_8$
are affected neither by constructing $e_4$ in cases 2 and 5 nor by increasing 
the capacity of $e_{23}$ in cases 3 and 6. This verifies the finding in Remark \ref{remarkfindings}-(i). 
\par 
Second, by comparing the weights and the 
variances in the lines in case 4 with those in case 5, it may be noted in Table \ref{table2} and \ref{table3} that both the weights and the variances in $e_5-e_{29}$ are not changed when $e_4$ is constructed in case 5. This is because these lines are not in the cycle-cluster that includes $e_4$. 
Similarly, by comparing the weights and the variances in the lines in case 5 with those in 
case 6, it is seen in Table \ref{table2} and \ref{table3}
that both the weights and the variances in $e_1-e_8$ are not influenced 
by increasing the capacity of $e_{23}$. 
This is due to the fact that these lines are not in the cycle-cluster that includes $e_{23}$. Thus, 
the findings in Remark \ref{remarkfindings}-(ii) is verified . 
\par 
Third, we evaluate the findings in Remark \ref{remarkfindings}-(iii). 
The effects of constructing $e_4$ have already been analyzed, where the variances in the lines in the cycle-cluster of $(e_1,e_2,e_3,e_4)$ all decrease while those in the other lines are not affected. 
When comparing the variances in the lines in case 2 with those in case 3 in Table \ref{table3},
it is found that the variances in $e_{11},e_{14},e_{17}-e_{29}$ all decrease after increasing the capacity of $e_{23}$ from $10$ to $20$. We remark that those in $e_9,e_{10},e_{12},e_{13}$ also 
decrease, which are not explicitly shown in the table because of the limited precision. This 
indicates that the variances of the lines in a cycle-clusters all decrease if the capacity
of a line in this cycle-cluster increases. However, in practice, constructing
new lines or increasing the capacity of lines also changes
the power flows, which further influence the weight $l_{c_{ij}}$. 
For example, when comparing the weights in case 5 with those in case 6, it is shown in Table \ref{table2} that after increasing 
the capacity of $e_{23}$ in case 6, the weights of $e_{24},e_{25},e_{26}$
decrease from $9.7528,9.9266,9.9978$ to $9.6934,9.8933,9.9896$ respectively.
We remark that similar as in case 3, the variances in $e_9,e_{10},e_{12},e_{15},e_{16}$
also decrease, which are not explicitly shown due to the limitation of the precision.
Although only some of the weights decrease, as shown in Table \ref{table3}, the variances in $e_9-e_{29}$ all decrease. This is due to the fact that the negative
impact brought by the decrease in the weights cannot
overcome the positive impact brought by increasing the capacity 
of $e_{23}$. However, if the negative impact surpasses the
positive impact, then the variance will increase, which may happen in a subset of networks. 
\par 
Finally, we verify the findings in Remark \ref{remarkfindings}-(iv). We focus on Cases 2-3 with $\delta_{ij}^*=0$ for all the lines that 
are not changed by either constructing new lines or increasing the line capacity. 
The cycle-cluster $\{e_1,e_2,e_3,e_4\}$ includes a cycle. 
The basis vector corresponding to this cycle is $\bm \xi_1=[-1,-1,-1,1,0,\cdots,0]^{\bm\top}$.
By scaling this vector to unit length, we obtain $\bm X_1=[-1/2,-1/2,-1/2,1/2,0,\cdots,0]^{\bm\top}$.
From formula (\ref{theoremmain-1}), we obtain that the diagonal 
elements $\bm Q_\delta$ at positions (1-4) are all $3/80$, which is consistent with the values shown in Table \ref{table3}. 
Hence, the construction of $e_4$ decreases
the variances of the phase angle differences, and the size of the decrease depends on the length 
of the cycle. It is verified that the variances in $e_1-e_4$ in cases 5 and 6 can also be calculated by (\ref{singlecycle}) for simplicity. 
Let us next focus on the
conservative estimation of the variances in the lines in a cycle-cluster by formula (\ref{singlecycle}). For example, the variance 
in $e_{29}$ in case 2 can be approximated as $0.0333$ for simplicity from 
formula (\ref{singlecycle}) by taking $\mathcal{E}_c=\{e_{22},e_{28},e_{29}\}$. 
This value is larger than $0.0326$ as shown in Table \ref{table3}. 
Because constructing new lines to form cycles or increasing the capacities 
of lines changes the power flows, which may decrease
the weights of the lines in the cycle-cluster or even destroy the synchronization, it is complicated 
to analyse how the variances of the lines of this cycle-cluster change. 
However, in a real network, the phase angle differences are usually small, and 
the weight $l_{c_{ij}}\approx l_{ij}$, which is often assumed in the 
investigation of the synchronization of power systems \citep{optimal_inertia_placement,H2norm}. In this case, the negative 
influences on the weight can be neglected and the variances decrease if 
new lines are constructed to form small cycles or the capacities of the lines are 
increased. The reduction in the variances can be approximated using (\ref{singlecycle}).
\par 
In regard to the bounds of the variance matrices for the networks with non-uniform
disturbance-damping ratio, it is shown for case 7 in Table \ref{table1} and \ref{table3} that the variances of the frequencies at the nodes and the phase angle differences in the 
lines are both constrained by the lower bound and the upper bound in (\ref{theorem4-1}) and (\ref{theorem4-2}) respectively.
For the frequency,  it is demonstrated that the variance at the node which possesses 
the largest disturbance-damping ratio is closer to the upper bound and that at the node with the smallest disturbance-damping ratio is closer to the lower bound.  For example,  the variance at node $5$, which has the largest disturbance-damping ratio $\eta_5=\sqrt{5}$, is $0.1901$ which is closer to the upper bound $\sqrt{5}/10=0.2236$. However,
those at the nodes $1, 6-23$ with the smallest disturbance-damping ratio are all closer to the lower bound $1/2$.
For the phase angle differences,  the variance in the lines which connect nodes with larger disturbance-damping ratio 
are usually larger.  For example, the variance in $e_8$ is $0.0904$ which becomes closer to the upper bound $0.1198$ compared with its value in case 6.  However,  the variances in the lines which are far away from the nodes with 
larger disturbance-damping ratio are closer to the lower bounds.  This is seen from the variance 
in lines $e_{26}-e_{29}$. 
\par 
In regard to the vulnerable nodes, it is found in Table \ref{table1} that nodes $1, 6-23$ in case 1-6 and 
node $6$ in case 7 are the most vulnerable nodes.  The remedy methods include increasing the inertia and decreasing 
the disturbance-damping ratio at these nodes or their neighbor nodes. 
With respect to the vulnerable lines, it is seen in Table \ref{table3} for cases 1-7 that the single lines 
are usually the vulnerable lines, for example, lines $e_5-e_8$. The remedy method includes increasing the capacities of these lines and 
constructing new lines to include these lines into cycles.


\section{Conclusion}\label{Section:conclusion}

In this paper, based on a stochastic Gaussian system, we have investigated the dependence of the fluctuations 
in a power system on system parameters when subjected to 
stochastic disturbances.
The dynamics of turbine-governors of the synchronous machines and that of voltage may be considered in the system \citep{passivityrobust}. By the method proposed in this paper, the impact 
of the system parameters on the fluctuations of the frequency and voltage at each node and the phase angle difference in each line can be investigated. 
In that case, the system parameters includes the ones in the dynamics of the turbine-governor and voltage besides 
those studied in this paper. 
\par 
A future investigation will address the deduction of explicit formulas for the 
variance matrices of the frequencies at the nodes and the phase 
angle differences in the lines in the network with non-uniform 
disturbance-damping ratio and lossy transmission lines. 
\par  


{\small
\bibliographystyle{dcu} 
\bibliography{autosam}
}   



\appendix
\section{Appendix}

\subsection{The variance matrix of a linear stochastic process}\label{appendix1}
\begin{definition}
Consider a linear stochastic system 
 \begin{align*}
 \text{\emph{d}}\bm x(t)&=\bm A\bm x(t)\text{\emph{d}}t+\bm B\text{\emph{d}}\bm \mu(t), \bm x(0)=\bm x_0,\\
 \bm y&=\bm C\bm x(t), 
 \end{align*}
where $\bm x\in\mathbb{R}^n$, $\bm A\in\mathbb{R}^{n\times n}$, $\bm B\in\mathbb{R}^{n\times m}$, $\bm x_0\in G(\bm 0,\bm Q_{x_0})$ is a Gaussian random variable where $\bm Q_{x_0}\in\mathbb{R}^{n\times n}$ is the variance matrix of $\bm x_0$, $\bm C\in\mathbb{R}^{z\times n}$, $\bm \mu\in\mathbb{R}^m$ is standard Brownian motion, $\bm y\in\mathbb{R}^z$ is the output. It follows
from \cite[Theorem 1.52]{linearOptimalSystems} that the state $\bm x$ and $\bm y$ are Gaussian process, i.e.,
for all $t>0$, 
\begin{align*}
\bm x(t)\in G(\bm 0,\bm Q_{x,tv}(t)),~~\bm y(t)\in G(\bm 0,\bm Q_{y,tv}(t))
\end{align*}
where the variance matrix $\bm Q_{x,tv}(t)\in\mathbb{R}^{n\times n}$ of $\bm x(t)$ is
\begin{align*}
\bm Q_{x,tv}(t)=\emph{e}^{\bm At}\bm Q_{x_0}\emph{e}^{\bm A^\top t}+\int_0^t{\emph{e}^{\bm A\tau}\bm B\bm B^{\bm\top}\emph{e}^{\bm A^{\bm\top}\tau}\text{\emph{d}}\tau}
\end{align*}
and the variance matrix $\bm Q_{y,tv}(t)\in\mathbb{R}^{z\times z}$ of $\bm y(t)$ satisfies $\bm Q_{y,tv}(t)=\bm C\bm Q_{x,tv}(t) \bm C^\top$. The matrix $\bm Q_{x,tv}(t)$ satisfies the matrix differential equation
\begin{subequations}
\begin{align}\label{matrixdiff}
\dot{\bm Q}_{x,tv}(t)&=\bm A\bm Q_{x,tv}(t)+\bm Q_{x,tv}(t)\bm A^\top+\bm B\bm B^\top,\\
\bm Q_{x,tv}(0)&=\bm Q_{x_0}.
\end{align}
\end{subequations}
In addition, if $\bm A$ is Hurwitz, then there exists an invariant distribution 
of the stochastic processes $\bm x(t)$ and $\bm y(t)$ with asymptotic variance matrices 
\begin{align*}
\bm Q_x=\lim_{t\rightarrow \infty}\bm Q_{x,tv}(t)=\int_0^{+\infty}{\emph{e}^{\bm A\tau}\bm B\bm B^{\bm\top}\emph{e}^{\bm A^{\bm\top}\tau}\text{\emph{d}}\tau},
\end{align*}
and $\bm Q_y=\bm C\bm Q_x\bm C^\top$. The matrix $\bm Q_x$, which is called the controllability Gramian of 
the pair $(\bm A,\bm B)$, is the unique solution of the Lyapunov equation due to the Hurwitz condition \citep{H2norm_book_toscano,H2norm_another_form}, 
 \begin{align}
  &\bm A\bm Q_x+\bm Q_x\bm A^{\bm\top}+\bm B\bm B^{\bm\top}=\bm 0.
 \end{align}
which can be either derived from the limit of the differential equation (\ref{matrixdiff}) or from
\begin{align*}
\bm A\bm Q_x+\bm Q_x\bm A^{\bm\top}&=\int_{0}^{+\infty}\frac{\emph{d}}{\emph dt}(\emph{e}^{\bm At}\bm B\bm B^{\bm\top}\emph{e}^{\bm A^{\bm\top}t})\emph{d}t\\
&=(\emph{e}^{\bm At}\bm B\bm B^{\bm\top}\emph{e}^{\bm A^{\bm\top}t})\Big|_0^{+\infty}=-\bm B\bm B^\top. 
\end{align*} 
\end{definition}
\subsection{The Moore-Penrose Pseudo inverse of real symmetric matrices}
\begin{theorem}\label{Lemma0}
Consider a real symmetric matrix $\bm S\in \mathbb{R}^{n\times n}$. There exists 
an orthogonal matrix $\bm V\in\mathbb{R}^{n\times n}$ such that
\begin{align*}
\bm V^{\bm\top} \bm S\bm V=\bm \Sigma
\end{align*}
where $\bm\Sigma=\text{diag}(\sigma_i)\in\mathbb{R}^{n\times n}$ is a diagonal matrix with the diagonal elements $\sigma_i$  being the eigenvalues of $\bm S$,  the column vectors $\bm v_i$ of $\bm V$ are orthonormal eigenvectors of $\bm S$ corresponding to the eigenvalue $\sigma_i$. In addition, the Moore-Penrose pseudo inverse is defined by the formula
\begin{align*}
\bm S^{\dag}=\bm V\Sigma^{\dag}\bm V^{\bm\top}=\sum_{i=1}^{n}\sigma_i^*\bm v_i\bm v_i^{\bm \top}
\end{align*}
where $\bm \Sigma^\dag=\text{diag}(\sigma_i^*)\in\mathbb{R}^{n\times n}$ with $\sigma_i^*=1/\sigma_i$ if $\sigma_i\neq 0$, otherwise $\sigma_i^*=0$ \citep{horn}. 
\end{theorem}

\subsection{The basis vectors of the kernel of $\widetilde{\bm C}\bm R^{1/2}$ }\label{kernelspace}

The cycle space of a graph is defined as the kernel 
of the incidence matrix $\widetilde{\bm C}$, which is a vector subspace in $\mathbb{R}^{m}$. 
By graph theory, we have $\text{rank}(\widetilde{\bm C})=n-1$. Hence, the dimension 
of the cycle space is $m-n+1$. It is obvious that the cycle space 
of an acyclic graph is an empty space. For a graph with cycles, 
the basis for the cycle space is derived by the following method: 
Considering a cycle $\mathcal{C}$ with a set $\mathcal{E}_c$ of edges in the graph $\mathcal{G}$,
we specify a direction for $\mathcal{C}$;
then, the vector $\bm\xi_c=[\xi_{c,1},\xi_{c,2},\cdots,\xi_{c,m}]^{\bm\top}\in\mathbb{R}^{m}$
such that 
\begin{eqnarray*}
\xi_{c,k}=
 \begin{cases}
         +1, &\text{\footnotesize if $e_k\in\mathcal{E}_c$ with direction $=$ the cycle direction},\\
         -1, &\text{\footnotesize if $e_k\in\mathcal{E}_c$ with direction $\neq$ the cycle direction},\\
         0,&\text{\footnotesize otherwise.}
        \end{cases}
\end{eqnarray*}
belongs to the kernel of $\widetilde{\bm C}$ such that $\widetilde{\bm C}\bm \xi_c=\bm 0$ \citep{NORMAN}.
The basis for the 
cycle space can be derived by taking
the vectors as $\bm \xi_c$ for $c=1,\cdots,m-n+1$ corresponding to the $(m-n+1)$~ \emph{fundamental
cycles} \cite[Theorem 1.9.6]{Diestel} in the graph. Because $\bm R$ is non-singular, the vectors $\bm R^{-1/2}\bm\xi_c$
for all the cycles are the basis vectors of the kernel of $\widetilde{\bm C}\bm R^{1/2}$. The orthonormal basis vectors $\bm X_i$ are obtained by Gram-Schmidt orthogonalization of the basis vectors $\bm R^{-1/2}\bm\xi_c$. 
The fundamental cycles can be obtained by the following method. 
Let $\mathcal{T}$ be a spanning tree of the graph $\mathcal{G}$. Then $\mathcal{T}$ has $n-1$ edges and 
there are $m-n+1$ edges of $\mathcal{G}$ lie outside of $\mathcal{T}$. Then for each of these $m-n+1$
edges $e\in\mathcal{E}\setminus\mathcal{E}(\mathcal{T})$, the graph $\mathcal{T}+e$ contains a cycle, which is a fundamental cycle. Note that the basis vectors of the cycle
space may not be unique due to the non-uniqueness of the spanning tree.

\setcounter{table}{0}
\renewcommand{\thetable}{\arabic{table}}
\onecolumn
\par 

\begin{table}[h]
\caption{Table describing the 7 cases in the example;
line $e_4$ is present in the network if the label is $1$
and not present if the label is $0$.}\label{table0}
\begin{tabular}{c|c|c|c|c|c|c|c} \hline
Case & Line  & Source  & Load  & \multicolumn{2}{c|}{ Line $l_{ij}$} & \multicolumn{2}{c}{  $b_i^2/d_i$}\\ \cline{5-8} 
     & $e_4$ & $P_i$  & $P_i$ & $e_{23}$  & others & 1-5&others \\ \hline
1    & 0     & 0      & ~ 0   & 10       & 10&1&1 \\ \hline
2    & 1     & 0      & ~ 0   & 10       & 10 &1&1 \\ \hline
3    & 1     & 0      & ~ 0   & 20       & 10&1&1 \\ \hline
4    & 0     & 3.6    & -1    & 10       & 10 &1&1\\ \hline
5    & 1     & 3.6    & -1    & 10       & 10 &1&1\\ \hline
6    & 1     & 3.6    & -1    & 20       & 10 &1&1\\ \hline
7    & 1     & 3.6    & -1    & 20       & 10 &$\sqrt{i}$&1\\ \hline
\end{tabular}
\end{table}

\par 
~~
\par 
\begin{table}[!ht]
\caption{The variances of the frequencies in the 7 cases in the example; 7L and 7U denotes
the lower and upper bounds in case 7; $\sqrt{5}\approx2.2236$. }
\label{table1}
\scalebox{0.86}{
\begin{tabular}{c |c| c| c| c| c |c|c|c|c|c|c|c} 
\hline
case & 1 & 2 & 3& 4 & 5 & 6 & 7& 8& 9 & 10& 11& 12 \\
\hline
 1-6,  7L& $1/2$ & $1/4$&$1/6$&$1/8$ &$1/10$&$1/2$&$1/2$&$1/2$&$1/2$&$1/2$&$1/2$&$1/2$ \\
\hline
7 & $0.5002$ & $0.3012$&$0.2600$&$0.2275$ &$0.1901$&$0.5107$&$0.5024$&$0.5054$&$0.5034$&$0.5002$&$0.5006$&$0.5038$\\
\hline
7U& $\sqrt{5}/2$ & $\sqrt{5}/4$&$\sqrt{5}/6$&$\sqrt{5}/8$ &$\sqrt{5}/10$&$\sqrt{5}/2$&$\sqrt{5}/2$&$\sqrt{5}/2$&$\sqrt{5}/2$&$\sqrt{5}/2$&$\sqrt{5}/2$&$\sqrt{5}/2$\\
\hline
 \hline
case & 13 & 14 & 15& 16 & 17 & 18& 19& 20& 21 & 22& 23&  \\
\hline
 1-6, 7L  & $1/2$&$1/2$&$1/2$ &$1/2$&$1/2$&$1/2$&$1/2$&$1/2$&$1/2$&$1/2$&$1/2$ \\
\hline
7 & $0.5025$& $0.5002$&$0.5000$&$0.5000$ &$0.5000$&$0.5000$&$0.5001$&$0.5004$&$0.5042$&$0.5003$&$0.5001$&\\
\hline
 7U& $\sqrt{5}/2$ & $\sqrt{5}/2$&$\sqrt{5}/2$&$\sqrt{5}/2$ &$\sqrt{5}/2$&$\sqrt{5}/2$&$\sqrt{5}/2$&$\sqrt{5}/2$&$\sqrt{5}/2$&$\sqrt{5}/2$&$\sqrt{5}/2$&\\\hline
\end{tabular}}
\end{table}

\par
~~
\par 
\begin{table}[!ht]
\caption{The weights of the lines of the network in Cases (4-7) in the example.}
\label{table2}
\scalebox{0.845}{
\begin{tabular}{c|c|c|c|c|c|c|c|c|c|c|c|c|c|c|c}
\hline
Case    & $e_1$ &$e_2$&$e_3$&$e_4$&$e_5$&$e_6$& $e_7$&$e_8$&$e_9$&$e_{10}$&$e_{11}$&$e_{12}$&$e_{13}$&$e_{14}$&$e_{15}$\\ \hline
   4 &$9.9499$ & $7.8460$ &$9.3295$ &$-$      &$9.9499$&$9.6561$&$9.8712$&$9.3295$&$9.9109$&$9.9945$&$9.8215$&$9.9193$&$9.7394$&$9.9549$&$9.9860$ \\ \hline
   5 &$9.8530$ & $9.3706$ &$9.9602$ &$9.6262$&$9.9499$&$9.6561$&$9.8712$&$9.3295$&$9.9109$&$9.9945$&$9.8215$&$9.9193$&$9.7394$&$9.9549$&$9.9860$\\ \hline 
   6  &$9.8530$ & $9.3706$ &$9.9602$ &$9.6262$&$9.9499$&$9.6561$&$9.8712$&$9.3295$&$9.9092$&$9.9941$&$9.8312$&$9.9209$&$9.7423$&$9.9607$&$9.9897$\\ \hline\hline
Case    & $e_{16}$ &$e_{17}$&$e_{18}$&$e_{19}$&$e_{20}$&$e_{21}$& $e_{22}$&$e_{23}$&$e_{24}$&$e_{25}$&$e_{26}$&$e_{27}$&$e_{28}$&$e_{29}$\\ \hline
   4 &$9.7337$ & $9.9540$ &$9.9086$ &$9.7745$&$9.6346$&$9.9380$&$9.8829$&$9.4709$&$9.7528$&$9.9266$&$9.9978$&$9.9687$&$9.9965$&$9.9198$\\  \hline
   5 &$9.7337$ & $9.9540$ &$9.9086$ &$9.7745$&$9.6346$&$9.9380$&$9.8829$&$9.4709$&$9.7528$&$9.9266$&$9.9978$&$9.9687$&$9.9965$&$9.9198$\\  \hline
   6-7 &$9.7430$ & $9.9434$ &$9.9271$ &$9.7973$&$9.6722$&$9.9495$&$9.9069$&$19.6990$&$9.6934$&$9.8933$&$9.9896$&$9.9852$&$9.9984$&$9.9300$\\ \hline   
\end{tabular}
}
\end{table}
\par
~~
\begin{table}[!ht]
\caption{The variances of the phase angle differences in the 7 Cases in the example; 7L and 7U denotes
the lower and upper bounds in case 7.} 
\label{table3}
\scalebox{0.84}{
\begin{tabular}{c|c|c|c|c|c|c|c|c|c|c|c|c|c|c|c}
\hline
Case    & $e_1$ &$e_2$&$e_3$&$e_4$&$e_5$&$e_6$& $e_7$&$e_8$&$e_9$&$e_{10}$&$e_{11}$&$e_{12}$&$e_{13}$&$e_{14}$&$e_{15}$\\ \hline
   1&$0.0500$ & $0.0500$ & $0.0500$ &$-$&$0.050$&$0.050$&$0.050$&$0.050$&$0.0394$&$0.0394$&$0.0298$&$0.0394$&$0.0394$&$0.0340$&$0.0281$ \\ \cline{1-16} 
   2&$0.0375$ & $0.0375$ & $0.0375$ &$0.0375$&$0.050$&$0.050$&$0.050$&$0.050$&$0.0394$&$0.0394$&$0.0298$&$0.0394$&$0.0394$&$0.0340$&$0.0281$ \\ \cline{1-16} 
   3&$0.0375$ & $0.0375$ & $0.0375$ &$0.0375$&$0.050$&$0.050$&$0.050$&$0.050$&$0.0394$&$0.0394$&$0.0297$&$0.0394$&$0.0394$&$0.0339$&$0.0281$ \\ \hline
   4 &$0.0503$ & $0.0637$ &$0.0536$ &$-$&$0.0503$&$0.0518$&$0.0507$&$0.0536$&$0.0397$&$0.0395$&$0.0302$&$0.0397$&$0.0403$&$0.0343$&$0.0283$ \\ \hline
   5 &$0.0383$ & $0.0396$ &$0.0380$ &$0.0389$&$0.0503$&$0.0518$&$0.0507$&$0.0536$&$0.0397$&$0.0395$&$0.0302$&$0.0397$&$0.0403$&$0.0343$&$0.0283$\\ \hline
   6,  7L &$0.0383$ & $0.0396$ &$0.0380$ &$0.0389$&$0.0503$&$0.0518$&$0.0507$&$0.0536$&$0.0397$&$0.0395$&$0.0301$&$0.0397$&$0.0402$&$0.0342$&$0.0283$\\ \hline
   7 &$0.0397$ & $0.0444$ &$0.0518$ &$0.0530$&$0.0551$&$0.0566$&$0.0524$&$0.0904$&$0.0398$&$0.0397$&$0.0304$&$0.0398$&$0.0403$&$0.0354$&$0.0290$\\ \hline
    7U &$0.0856$ & $0.0884$ &$$0.0849$$ &$0.0869$&$0.1124$&$0.1158$&$0.1133$&$0.1198$&$0.0889$&$0.0883$&$0.0674$&$0.0888$&$0.0900$&$0.0765$&$0.0632$\\ \hline 
    \hline 
Case    & $e_{16}$ &$e_{17}$&$e_{18}$&$e_{19}$&$e_{20}$&$e_{21}$& $e_{22}$&$e_{23}$&$e_{24}$&$e_{25}$&$e_{26}$&$e_{27}$&$e_{28}$&$e_{29}$\\ \hline
   1&$0.0262$ & $0.0304$ & $0.0292$ &$0.0352$&$0.0343$&$0.0352$&$0.0302$&$0.0430$&$0.0430$&$0.0430$&$0.0430$&$0.0430$&$0.0326$&$0.0326$ \\ \hline 
   2&$0.0262$ & $0.0304$ & $0.0292$ &$0.0352$&$0.0343$&$0.0352$&$0.0302$&$0.0430$&$0.0430$&$0.0430$&$0.0430$&$0.0430$&$0.0326$&$0.0326$ \\ \hline 
   3&$0.0262$ & $0.0303$ & $0.0291$ &$0.0351$&$0.0341$&$0.0351$&$0.0300$&$0.0231$&$0.0424$&$0.0424$&$0.0424$&$0.0424$&$0.0325$&$0.0325$ \\ \hline
   4 &$0.0267$ & $0.0306$ &$0.0296$ &$0.0359$&$0.0353$&$0.0356$&$0.0305$&$0.0451$&$0.0440$&$0.0434$&$0.0431$&$0.0432$&$0.0327$&$0.0328$ \\ \hline
   5 &$0.0267$ & $0.0306$ &$0.0296$ &$0.0359$&$0.0353$&$0.0356$&$0.0305$&$0.0451$&$0.0440$&$0.0434$&$0.0431$&$0.0432$&$0.0327$&$0.0328$\\ \hline
   6,  7L&$0.0267$ & $0.0305$ &$0.0293$ &$0.0357$&$0.0350$&$0.0354$&$0.0302$&$0.0235$&$0.0436$&$0.0429$&$0.0426$&$0.0426$&$0.0326$&$0.0327$\\ \hline  
      7&$0.0272$ &$0.0400$ &$0.0383$&$0.0365$&$0.0352$&$0.0356$&$0.0303$&$0.0318$&$0.0456$&$0.0432$&$0.0426$&$0.0426$&$0.0326$&$0.0328$\\ \hline   
      7U& $0.0597$ &$0.0683$ &$0.0656$&$0.0799$&$0.0783$&$0.0792$&$0.0676$&$0.0525$&$0.0976$&$0.0960$&$0.0952$&$0.0952$&$0.0729$&$0.0731$\\ \hline    
\end{tabular}
}
\end{table}
\twocolumn

\end{document}